\documentclass[%
superscriptaddress,
 amsmath,amssymb,
 superscriptaddress,
 aps,
 pre,
 twocolumn
]{revtex4-2}
\usepackage{graphicx}%
\usepackage{xcolor}
\usepackage{blkarray}%
\usepackage{mathtools}
\usepackage{comment} 

\newcommand{\tildeb}[1]{\stackrel{\sim}{\smash{#1}\rule{0pt}{1.25ex}}}
\begin{document}

\title{Particle-scale origin of quadrupolar non-affine displacement fields in granular solids}
\author{Evan P. Willmarth}
\affiliation{Department of Mechanical Engineering, Yale University, New Haven, Connecticut, 06520, USA}
\author{Weiwei Jin}
\thanks{Corresponding author: Weiwei Jin, microwei.jin@gmail.com}
\affiliation{Department of Mechanical Engineering, Yale University, New Haven, Connecticut, 06520, USA}
\author{Dong Wang}
\affiliation{Department of Mechanical Engineering, Yale University, New Haven, Connecticut, 06520, USA}
\author{Amit Datye}
\affiliation{Department of Mechanical Engineering, Yale University, New Haven, Connecticut, 06520, USA}
\author{Udo D. Schwarz}
\affiliation{Department of Mechanical Engineering, Yale University, New Haven, Connecticut, 06520, USA}
\author{Mark D. Shattuck}
\affiliation{Benjamin Levich Institute and Physics Department, The City College of The City University of New York, New York, New York 10031, USA}
\author{Corey S. O'Hern}
\affiliation{Department of Mechanical Engineering, Yale University, New Haven, Connecticut, 06520, USA}
\affiliation{Department of Applied Physics, Yale University, New Haven, Connecticut, 06520, USA}
\affiliation{Department of Materials Science, Yale University, New Haven, Connecticut, 06520, USA}
\affiliation{Department of Physics, Yale University, New Haven, Connecticut, 06520, USA}

\begin{abstract}
We identify the local structural defects that control the non-affine displacement fields in jammed disk packings subjected to athermal, quasistatic (AQS) simple shear. While complex non-affine displacement fields typically occur during simple shear, isolated effective quadrupoles are also observed and their probability increases with increasing pressure. We show that the emergence of an isolated effective quadrupole requires the breaking of an interparticle contact that is aligned with low-frequency, spatially extended vibrational modes. Since the Eshelby inhomogeneity problem gives rise to quadrupolar displacement fields in continuum materials, we reformulate and implement Eshelby's equivalent inclusion method (EIM) for jammed disk packings. Using EIM, we show that we can reconstruct the non-affine displacement fields for jammed disk packings in response to applied shear as a sum of discrete Eshelby-like defects that are caused by mismatches in the local stiffnesses of triangles formed from Delaunay triangulation of the disk centers.
\end{abstract}

\maketitle

\section{Introduction}

The response of disordered particulate systems to applied simple shear is spatiotemporally complex. The shear stress versus shear strain includes quasi-linear elastic segments of differing lengths that are punctuated by shear stress drops over a range of sizes, which indicate particle rearrangement events~\cite{maloney06, jin21, sopu21, zhang21, xu17, baggioli21,jin23}.
Experimental and numerical studies have also shown that collective, quadrupolar displacement fields occur during quasistatic simple shear~\cite{kang2023, maloney04, maloney06, jin21,sopu21,zhang21, dasgupta12, desmarchelier24}. This result is noteworthy since quadrupolar displacement fields are solutions to the Eshelby inhomogeneity problem for continuum materials with inclusions that have different elastic constants than the surrounding matrix~\cite{eshelby1957}. This connection highlights the possible important role of the Eshelby solutions in describing localized plastic rearrangement events even in disordered, particulate systems~\cite{dasgupta12, Sopu2017, mcnamara2016eshelby, NICOLAS2015}.
Currently, we do not know the frequency with which quadrupolar displacement fields occur in sheared particulate systems as a function of shear strain and packing fraction. For example, do quadrupolar displacement fields occur uniformly over the quasi-elastic strain segments or do they occur more frequently near large shear stress drops?

The mechanical response of crystalline solids is determined by the strength and location of topological defects, such as point defects and dislocations~\cite{lemaitre1994mechanics}. However, identifying the structural defects in disordered particulate systems that control the mechanical response is more challenging. Motivated by early experimental studies on bubble rafts, Argon coined the term shear transformation zones (STZs) to describe regions of large non-affine particle motion in disordered materials~\cite{argon1979plastic}. Falk and Langer pictured the STZs as structural defects in disordered solids that generate large non-affine particle displacements~\cite{falk1998dynamics}. However, currently there is no quantitative framework that directly links localized structural defects in disordered particulate solids to quadrupolar-like and other coherent structures in non-affine displacement fields~\cite{csopu2023stz}.

In addition, numerous studies have shown that ``soft spots'', or quasi-localized modes of the dynamical matrix with small eigenvalues, influence the mechanical response of amorphous solids~\cite{richard2021brittle,manning2011vibrational}.  These modes are often quadrupolar-like in structure~\cite{maloney06,silbert2005vibrations} and are frequently excited during applied shear~\cite{manning2011vibrational, ding2014soft}.  However, focusing on modes of the dynamical matrix de-emphasizes the role of local structural defects in determining the non-affine displacement fields. In particular, the connection between local structural defects and the activation of ``soft spots'' is missing~\cite{pan2017correlation}.  For example, it is not known what type of localized structural defect in a disordered particulate solid generates a single Eshelby-like quadrupolar displacement field. 

In this article, we develop a novel method to calculate the non-affine displacement fields of model granular solids (i.e., frictionless, bidisperse disks that interact via purely repulsive spring forces) in response to athermal, quasistatic simple shear in terms of Eshelby-like triangle ``defects''.  These defects represent the differences in the stiffness matrix of each triangle in the jammed disk packing relative to that from a reference unstressed network generated from Delaunay triangulation of the disk centers. In general, the stiffness matrix of a triangle can differ from that for the reference network due to missing interparticle contacts and pre-stress. A benefit of this new approach is that it emphasizes that structural defects in disordered particulate materials {\it cause} the non-affine displacement fields. Each triangle defect contributes a single quadrupolar-like displacement field (with a given amplitude and orientation) to the total displacement field in response to the applied simple shear strain. Thus, we can sum the contributions from each triangle defect (including interactions between them) to reconstruct the total non-affine displacement field after each simple shear strain increment is applied.

This article presents several important results. First, we show that the non-affine displacement fields of jammed bidisperse disk packings in response to athermal, quasistatic simple shear can be fit to the sum of effective quadrupolar displacement fields using solutions of the Eshelby inclusion problem for continuum materials~\cite{dasgupta12}. At low pressures near jamming onset, isolated effective quadrupoles {\it do not occur}. However, as the pressure increases, the probability for a few isolated effective quadrupoles to occur increases, and grows rapidly when the fraction of missing contacts $N_m/(N+1) \lesssim 0.05$, where $N_m=3N-N_c$ and $N_c$ is the number of distinct interparticle contacts among $N$ particles.
We find that an isolated effective quadrupole only occurs after breaking an interparticle contact that is aligned with the low-frequency modes of the dynamical matrix before the applied deformation, and the center of the quadrupole is located near the broken contact. Isolated effective quadrupoles do not occur when interparticle contacts form.  Moreover, we demonstrate that if we artificially ``heal'' pre-existing missing contacts (relative to the reference Delaunay-triangulated network) near the quadrupole center, the effective quadrupolar displacement field will not be activated by the applied simple shear. 

The remainder of the article is organized as follows. In Sec.~\ref{method}, we describe the numerical techniques for generating jammed bidisperse disk packings over a range of pressures, applying athermal, quasistatic simple shear to the jammed disk packings, and quantifying the non-affine displacement fields in response to the applied simple shear. In Sec.~\ref{results}, we present the main results including: (a) The probability that isolated effective quadrupoles occur during the athermal, quasistatic simple shear as a function of pressure; (b) A reformulation of the Eshelby inclusion and inhomogeneity problems for a single structural defect in particulate systems; (c) The solution of the multiple Eshelby inclusions and inhomogeneity problems for particulate systems; and (d) The identification of the specific interparticle contacts that when broken during applied shear will generate displacement fields with a single effective quadrupole. A summary of the particle-scale origins of single effective quadrupolar displacement fields and promising future research directions, such as extending the Eshelby inclusion and inhomogeneity problems to discrete systems with long-range attractive interactions and in three dimensions, are presented in Sec.~\ref{summary}. 
We also include five Appendices. In Appendix~\ref{appendix:A}, we provide a brief summary of the Eshelby inclusion and inhomogeneity problems for both continuum and discrete particulate systems. In Appendix~\ref{appendix:B}, we calculate the stiffness tensor for a single Delaunay triangle in a jammed disk packing. In Appendix~\ref{appendix:C}, we define the global stiffness matrix in terms of the local triangle stiffness tensors. Appendix~\ref{appendix:D} describes the parameter $\chi$, which quantifies the change in the lowest eigenvalue of the dynamical matrix of jammed disk packings before and after a change in the interparticle contact network.  Appendix~\ref{appendix:E} shows the distribution of the participation ratio $\rho$ of the eigenvectors ${\vec e}_k$ of the dynamical matrix prior to a contact break and illustrates how $\rho({\vec e}_k)$ changes before and after the contact break. 

\section{Methods}
\label{method}

The methods section is divided into two subsections. In Sec.~\ref{method:A}, we describe the numerical techniques for generating jammed bidisperse disk packings as a function of pressure and deforming them using athermal, quasistatic simple shear.  In Sec.~\ref{method:B}, we describe our methods for quantifying the non-affine displacement fields of the jammed disk packings following each simple shear strain increment.  In particular, we fit the non-affine displacement fields to a linear combination of the fields generated from Eshelby inclusions in an elastic matrix with specified strengths, locations, and orientations.

\subsection{Generation and athermal, quasistatic simple shear of jammed disk packings}
\label{method:A}

We investigate the mechanical response of jammed packings of $N$ frictionless bidisperse disks (with the same mass $m$) in two dimensions. The disks interact via the pairwise, purely repulsive linear spring potential,
\begin{equation}
    \label{potential}
    U_{ij}(r_{ij}) = \frac{k_{ij}}{2} \left( \sigma_{ij} - r_{ij} \right)^{2}
    \Theta \left( 1- r_{ij}/\sigma_{ij}  \right),
\end{equation}
where $r_{ij}$ is the center-to-center distance between disks $i$ and $j$, $\sigma_{ij} = (\sigma_i + \sigma_j)/2$ is the average of their diameters $\sigma_i$ and $\sigma_j$, and $\Theta(\cdot)$ is the Heaviside step function that prevents interactions between non-contacting disks. For most cases, the spring constant $k_{ij}$ equals a fixed value $k$ for overlapping disk pairs and is zero otherwise.

The packings consist of bidisperse mixtures with equal numbers of large and small disks, with a diameter ratio of $\sigma_{l}/\sigma_{s}=1.4$. This bidisperse particle size distribution is widely used in both simulations and  experiments~\cite{puckett2013equilibrating,wang2022experimental,jiang2025experimental} to prevent crystallization~\cite{koeze2016mapping}. We use $m$, $\sigma_s$, $k$, and $k \sigma_s^2$ as the units for mass, length, stress, and energy, respectively. To generate jammed disk packings, we randomly place $N/2$ large and $N/2$ small disks in a parallellogram-shaped box with side lengths $\mathcal{L}_x$ and $\mathcal{L}_y$ and periodic boundary conditions in the $x$- and $y$-directions at an {\it initial} packing fraction $\phi_0 =N\pi(\sigma_l^2+\sigma_s^2)/(8\mathcal{L}_x\mathcal{L}_y) =0.1$. (Note that this expression for the packing fraction double-counts the areas of overlap between disks in packings that are overcompressed.) To obtain a jammed packing at a given pressure, we perform a sequence of compression steps $\Delta \phi$ (starting with the random disk configuration at $\phi_0=0.1$) with each step followed by minimization~\cite{bitzek2006} of the total potential energy $U = \sum_{i>j} U_{ij}(r_{ij})$ until the pressure $p=(\Sigma_{xx}+\Sigma_{yy})/2$
satisfies $|p/p_0-1|<10^{-4}$, where $p_0$ is the target pressure, 
\begin{equation}
\label{pressure}
\Sigma_{\alpha \beta}=\frac{1}{\mathcal{L}_x\mathcal{L}_y} \sum_{i>j} r_{ij\alpha} f_{ij\beta}
\end{equation}
is the virial stress tensor, $r_{ij\alpha}$ is the $\alpha$-component of the separation vector ${\vec r}_{ij}$, and ${f}_{ij\beta}$ is the $\beta$-component of the interparticle force, ${\vec f}_{ij} = -(dU_{ij}/dr_{ij}) {\hat r}_{ij}$. We first generate jammed disk packings at low pressure $p=10^{-6}$ (with a packing fraction $\phi\sim0.84$) and then successively compress each packing to a target pressure that spans the range $10^{-6} \le p_0 \le 0.32$. For all packings over this range of pressure, we apply athermal, quasistatic simple shear (where $x$ is the shear direction and $y$ is the shear-gradient direction) with a strain increment of $\Delta\gamma=2\times 10^{-5}$ to a total strain of $\gamma=0.2$. At each strain increment, the simple shear deformation 
\begin{equation}
    \label{eq:eps_G}
    \boldsymbol{\epsilon}^A = \begin{bmatrix}
        0 & \Delta\gamma \\
        0 & 0
    \end{bmatrix}
\end{equation}
is first applied to both the boundary of the simulation box and the disk positions, i.e., $\vec{a}_1^A=(\boldsymbol{\epsilon}^A+\boldsymbol{I})\vec{a}_1$, $\vec{a}_2^A=(\boldsymbol{\epsilon}^A+\boldsymbol{I})\vec{a}_2$, and $\vec{r}_i^A=(\boldsymbol{\epsilon}^A+\boldsymbol{I})\vec{r}_i$, where $\boldsymbol{I}$ is the $2\times2$ identity matrix, $\vec{a}_1=[\mathcal{L}_x,0]^T$ and $\vec{a}_2=[\gamma \mathcal{L}_y,\mathcal{L}_y]^T$ are the vectors that define the boundary vertexes at strain $\gamma$, and $\vec{r}_i$ is the position of disk $i$.
Then, minimization of the total potential energy of the disk packing is performed, while the box boundaries are fixed. Since the deformation protocol is strain-controlled, there is no work associated with the boundary motion. Thus, the total energy of the system is the elastic strain energy given by the total potential energy $U$.

\subsection{Characterization of the non-affine displacement fields}
\label{method:B}

We define the non-affine displacement of disk $i$ in a jammed disk packing undergoing athermal, quasistatic simple shear (where $x$ is the shear direction and $y$ is the shear-gradient direction) as
\begin{equation}
    \begin{bmatrix}
        u_{ix}\\
        u_{iy}
    \end{bmatrix}
    =
    \begin{bmatrix}
        r^\prime_{ix}-r_{ix}-r_{iy}\Delta\gamma\\
        r^\prime_{iy}-r_{iy}
    \end{bmatrix},
\end{equation}
\noindent where $[r_{ix}, r_{iy}]^T$ and $[r^\prime_{ix}, r^\prime_{iy}]^T$ are the positions of disk $i$ before and after an applied strain increment $\Delta\gamma$ (followed by minimization of the total potential energy), subject to shear periodic boundary conditions. The non-affine displacement field for the packing is given by $\vec{u} = [u_{1x}, u_{1y}, u_{2x}, u_{2y}, ...,u_{Nx}, u_{Ny}]^T$.

In general, the non-affine displacement fields of jammed disk packings in response to applied simple shear are disordered. Below, we investigate the probability that isolated quadrupolar non-affine displacement fields occur in jammed disk packings during simple shear as a function of pressure. We determine when isolated effective quadrupoles occur by fitting the non-affine displacement field in response to a small increment of simple shear strain to a linear superposition of the solutions of Eshelby inclusions under pure shear eigenstrains within an infinite, homogeneous elastic matrix with plane strain conditions~\cite{dasgupta12},
\begin{subequations}
\begin{align}
u_x(\vec{r}) &= \frac{\epsilon_0 a^2}{4(1 - \nu) r^2} \bigg[\left( 2 - 4\nu + \frac{a^2}{r^2} \right) (x \cos 2\alpha + y \sin 2\alpha) \nonumber \\
&\quad + \bigg( 1 - \frac{a^2}{r^2} \bigg) \frac{(x^2 - y^2) \cos 2\alpha + 2xy \sin 2\alpha}{r^2} 2x \bigg],\\
u_y(\vec{r}) &= \frac{\epsilon_0 a^2}{4(1 - \nu) r^2} \bigg[\left( 2 - 4\nu + \frac{a^2}{r^2} \right) (x \sin 2\alpha - y \cos 2\alpha) \nonumber \\
&\quad  + \bigg( 1 - \frac{a^2}{r^2} \bigg) \frac{(x^2 - y^2) \cos 2\alpha + 2xy \sin 2\alpha}{r^2} 2y \bigg],
\end{align}
\label{eq:circEsh}
\end{subequations}
where $\epsilon_0$ is the overall magnitude of the strain, $\nu$ is the Poisson's ratio of the surrounding elastic matrix, $a$ is the radius of the circular inclusion, $\vec{r}=[x, y]^T$ is the position in the packing relative to the center of the inclusion, and $\alpha$ is the orientation of the quadrupole. To account for periodic boundary conditions, we compute the non-affine displacement of each particle $i$ induced by an inclusion $q$ summed over all eight periodically replicated images in addition to the disk packing in the central box:
\begin{equation}
\label{fiteq}
    \begin{bmatrix}
u_{ix}\\
u_{iy}
\end{bmatrix}_q
=
\sum^8_{k=0}\begin{bmatrix}
u_x(\vec{r}^{(k)}_i)\\
u_y(\vec{r}^{(k)}_i)
\end{bmatrix}_q
\Theta\left(\frac{3}{2}\sqrt{\mathcal{L}_x\mathcal{L}_y}-
\left\lVert\vec{r}^{(k)}_i\right\rVert
\right),
\end{equation}
where the superscript $(k)$ refers to the central box $(0)$ and its periodically replicated images $(1), (2),..,(8)$, $\vec{r}^{(k)}_i$ denotes the position relative to the center of the inclusion in the central box, and $\|{\vec X} \|$ indicates the magnitude of the vector ${\vec X}$.  The cutoff of $3\sqrt{\mathcal{L}_x\mathcal{L}_y}/2$ in the distance from the center of the quadrupole was selected to maximize the agreement between the non-affine displacement field generated from a single quadrupole in Eq.~\ref{eq:circEsh} summed over the central and image boxes and the results from athermal, quasistatic simple shear. 
Assuming that $n_{\rm eff}$ inclusions are located in the central box and they interact linearly, the total inclusion-induced non-affine displacement field of each particle $i$ is
\begin{equation}
    \begin{bmatrix}
u_{ix}\\
u_{iy}
\end{bmatrix}_{T}
=
\sum^{n_{\rm eff}}_{q=1}\begin{bmatrix}
u_{ix}\\
u_{iy}
\end{bmatrix}_{q}.
\label{eq:eshT}
\end{equation}

To quantify the number of effective quadrupoles in a given displacement field, we initially place a large number of quadrupoles (Eq.~\ref{eq:circEsh}) at the positions of the negative topological charges~\cite{vaibhav2025} of the non-affine displacement field after a single strain step $\Delta \gamma$ of athermal, quasistatic simple shear. 
Other quantities, such as the squared non-affine displacement $D^2_{\rm min}$~\cite{falk1998dynamics}, the quadrupolar charge $Q$~\cite{mondal2023dipole}, and the Burgers ring~\cite{bera2025burgers}, can also be used to identify the centers of quadrupoles.
The initial fit aligns the non-affine displacement field from the simulations to that in Eq.~\ref{eq:eshT} with many quadrupoles. The quadrupoles with the largest $\|\vec{u}_{q}\|$ are then used as initial conditions for fitting the non-affine displacement field of the simulation to a small number $n_{\rm eff}$ of {\it effective} quadrupoles. The quality of the effective quadrupolar representation is assessed using the coefficient of determination $0 < R^2(n_{\rm eff}) < 1$. 

\begin{figure}
    \centering
    \includegraphics[width=\linewidth]{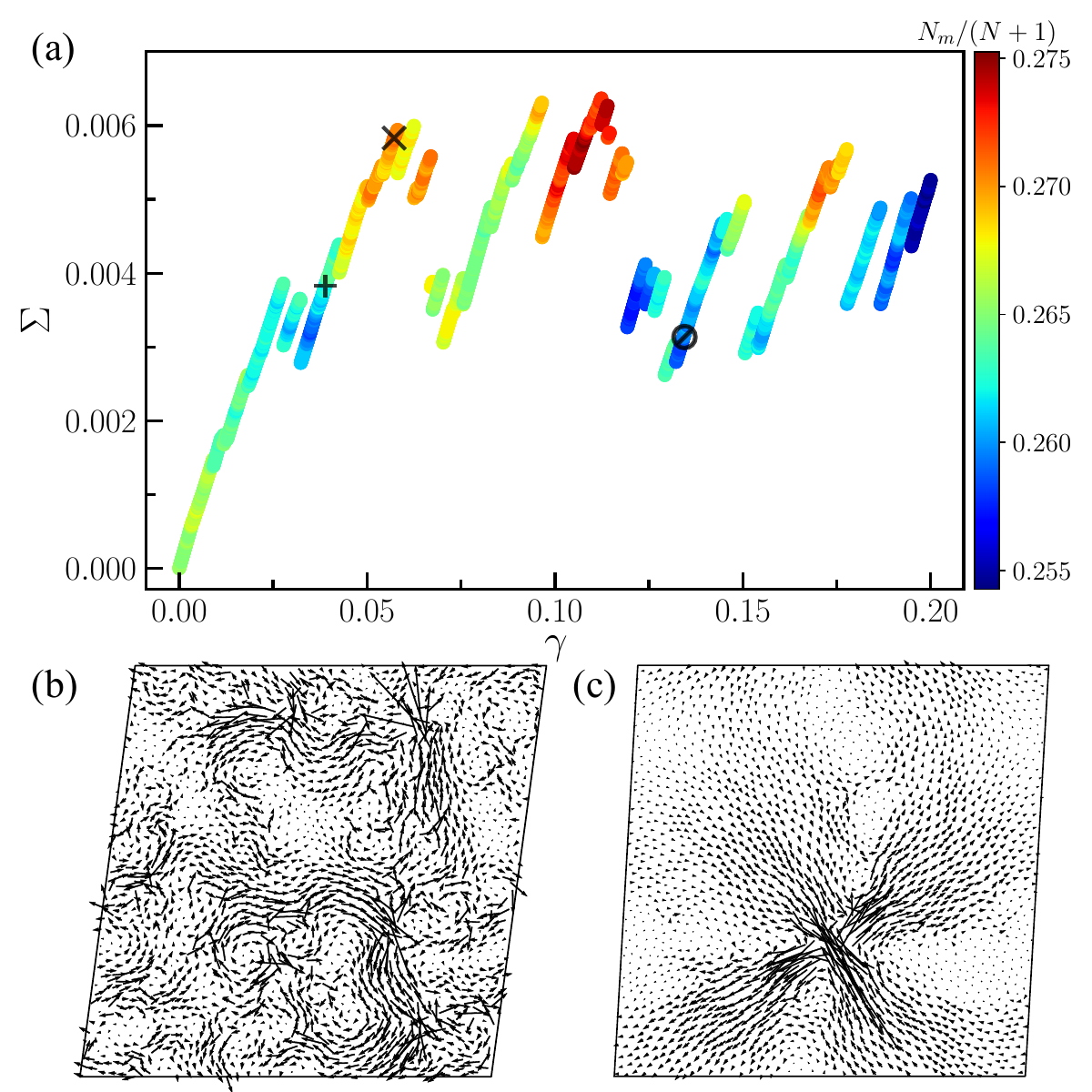}
    \caption{(a) Shear stress $\Sigma$ plotted versus shear strain $\gamma$ for a bidisperse mixture (diameter ratio $\sigma_l/\sigma_s=1.4$) of $N = 2048$ disks at an initial pressure of $p = 0.1$ ($\phi=1.0$) undergoing athermal, quasistatic simple shear with strain increment $\Delta \gamma = 2 \times 10^{-5}$. The data is shaded from violet to dark red as the fraction of missing contacts, $N_m/(N+1)$, increases. The black arrows in panels (b) and (c) show the corresponding non-affine displacement fields, $\vec{u}_b$ and $\vec{u}_c$, at the strains marked by the circle and $\times$, respectively, in (a). The $\times$ and $+$ signs in panel (a) indicate the strain of the non-affine displacement fields in Fig.~\ref{fig:fit} (a) and (b). $\vec{u}_b$ and ${\vec u}_c$ are magnified by a factor of 40000 and 5000, respectively, to improve visualization.}
    \label{fig:ss}
\end{figure}

\section{Results}
\label{results}
In this section, we characterize the non-affine displacement fields for jammed disk packings that undergo athermal, quasistatic simple shear. In Sec.~\ref{results:A}, we first determine the probability that isolated effective quadrupoles occur as a function of pressure, and show that the probability increases strongly with pressure. To understand the formation of single effective quadrupoles, in Sec.~\ref{results:B}, we introduce the Eshelby inclusion and inhomogeneity problems for continuum materials and then reformulate them for particulate systems, such as jammed disk packings. In particular, we model each triangle in the Delaunay triangulation of the disk packing as a discrete Eshelby inclusion, where triangle stiffness mismatches induce localized non-affine displacements.
In Sec.~\ref{results:C}, we describe a novel methodology for calculating the eigenstrains of individual Delaunay triangles that are used to reconstruct the non-affine displacement fields following a small shear strain increment. Our analysis reveals that triangles with large eigenstrains are concentrated near the centers of isolated effective quadrupoles and that clusters of missing contacts (compared to the Delaunay triangulation) near the centers of effective quadrupoles control their formation and stability. Finally, in Sec.~\ref{results:D}, we determine the necessary conditions for the activation and dissolution of isolated effective quadrupoles in the displacement field, which involve the breaking and formation of particle contacts near the center of the quadrupole.

\begin{figure}[t]
    \centering
    \includegraphics[width=0.75\linewidth]{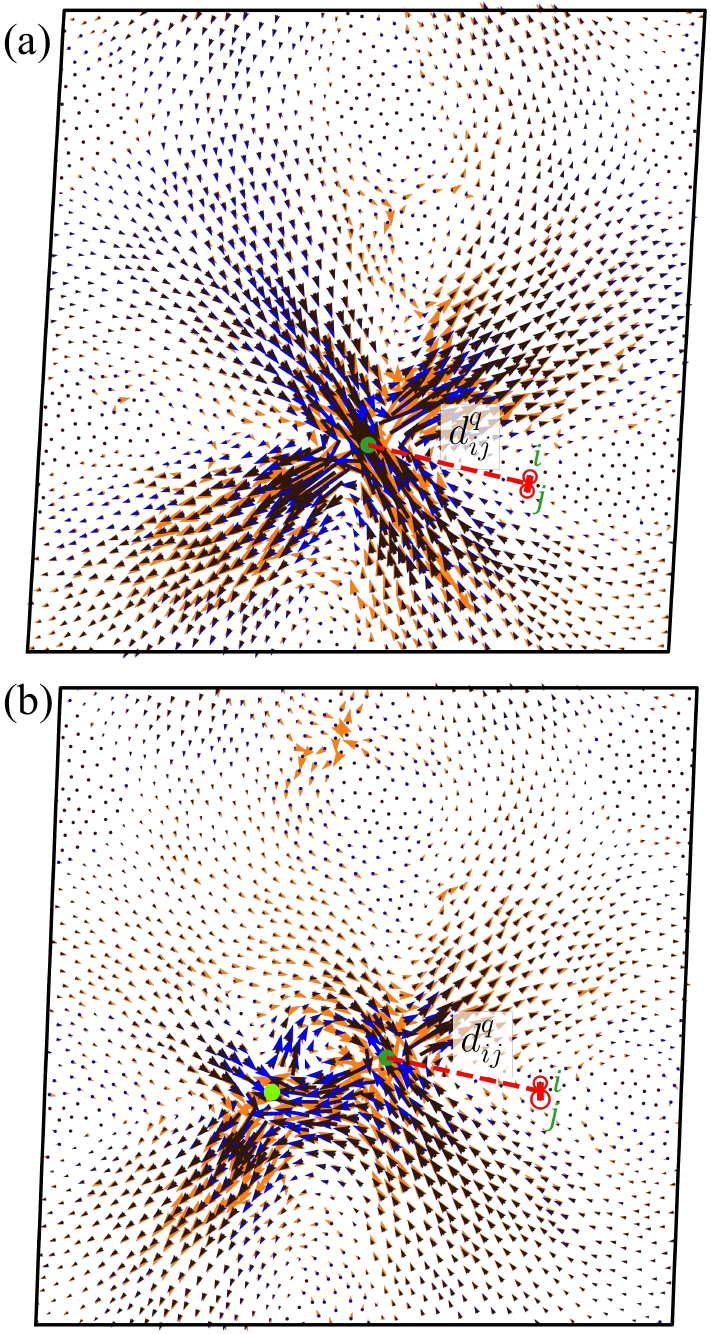}
    \caption{Non-affine displacement fields (orange-filled arrows) from the simulations of athermal, quasistatic simple shear in Fig.~\ref{fig:ss}: (a) at $\gamma=0.057$ fit to a single effective quadrupole with $R^2(1) = 0.87$ and (b) at $\gamma=0.039$ fit to two effective quadrupolar structures with $R^2(2) = 0.79$. The fitted quadrupolar non-affine displacement fields are represented by blue-filled arrows, with their centers marked by filled green circles. When the simulation and fitted non-affine displacement fields overlap, the arrows are shaed black. In (b), the quadrupole on the right (darker green) and the one on the left (lighter green) contribute $55\%$ and $45\%$ to the total non-affine displacement field, respectively. The dashed lines in both panels define the distance $d_{ij}^q$ from the center of quadrupole $q$ to the center of the bond between disks $i$ and $j$.}
    \label{fig:fit}
\end{figure}

\subsection{Shear stress versus shear strain}
\label{results:A}

In Fig.~\ref{fig:ss} (a), we show the shear stress $\Sigma=-\Sigma_{xy}$ versus shear strain $\gamma$ curve for jammed disk packings undergoing athermal, quasistatic simple shear.  $\Sigma(\gamma)$ possesses quasi-linear elastic segments punctuated by sudden stress drops~\cite{jin21}. In addition, the interparticle contact network changes frequently during simple shear. In this work, we focus on the mechanical response of jammed disk packings within each quasi-elastic segment, not during the stress drops, where the particle displacement fields evolve significantly. At most strain values, the non-affine displacement field ${\vec u}$ is disordered with no large-scale coherent structures (e.g. Fig.~\ref{fig:ss} (b)), whereas at particular strains, a single isolated quadrupole can form as shown in Fig.~\ref{fig:ss} (c). To quantify the probability with which isolated effective quadrupoles occur in jammed disk packings during athermal, quasistatic simple shear, we fit ${\vec u}$ at each $\gamma$ to a linear superposition of Eshelby-inclusion-induced quadrupoles, as described in Sec.~\ref{method:B}. In Fig.~\ref{fig:fit} (a), we show a non-affine displacement field that is well-described by a {\it single} effective quadrupole $n_{\rm eff}=1$, while the non-affine displacement field in Fig.~\ref{fig:fit} (b) is poorly fit by a single effective quadrupole with $R^2(1)<0.40$. Instead, it is better described by the sum of two effective quadrupoles ($n_{\rm eff}=2$) that contribute $45\%$ and $55\%$ to the total non-affine displacement field, respectively, with $R^2(2)=0.79$. We classify the fits to a superposition of quadrupoles as successful when they satisfy $R^2(n_{\rm eff})>R^2_{\rm cut}$, where $R^2_{\rm cut}=0.7$.

\begin{figure}[t]
    \centering
    \includegraphics[width=\linewidth]{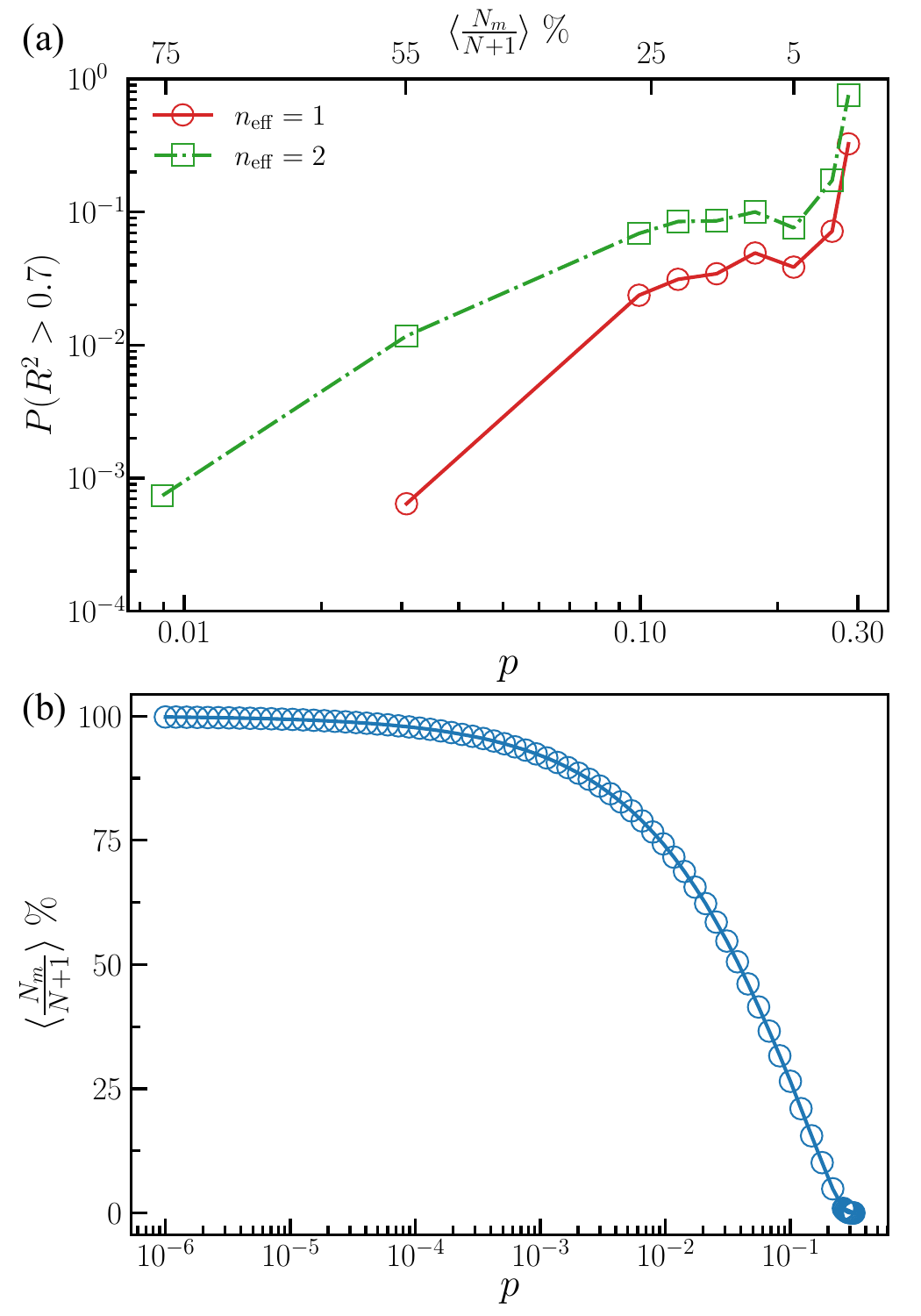}
    \caption{(a) Probability that the fitted non-affine displacement field ${\vec u}$ possesses $R^2(n_{\rm eff}) > 0.7$ plotted as a function of pressure $p$. The data is obtained by fitting ${\vec u}$ at the beginning of each quasi-elastic segment of shear stress versus strain to the non-affine displacement field generated by either one ($n_{\rm eff} = 1$, circles) or two ($n_{\rm eff} = 2$, squares) effective quadrupoles. (See Eq.~\ref{fiteq}.) The average fractional number of missing contacts $\langle N_m/(N+1)\rangle$ at each $p$ is displayed on the top axis. (b) $\langle N_m/(N+1)\rangle$ plotted as a function of $p$ for the same data in (a).}
    \label{fig:probQuad}
\end{figure}

In Fig.~\ref{fig:probQuad} (a), we determine the probability $P(R^2 > 0.7)$ that the non-affine displacement fields during athermal, quasistatic simple shear can be successfully fit to either one or the sum of two effective quadrupoles. Note that the non-affine displacement fields are similar for jammed disk packings within the same geometrical family (i.e., disk packings at different $\gamma$, but they possess the same interparticle contact network~\cite{tuckman2020}). Thus, in Fig.~\ref{fig:probQuad} (a), we only present results for $P(R^2 > 0.7)$ for disk packings at the beginning of each geometrical family over the full range of $\gamma$. In general, $P(R^2 > 0.7)$ increases with increasing pressure, but there is a sharp increase in $P(R^2 > 0.7)$ for $p \gtrsim 0.2$. 

To understand the structural origin of the coherent non-affine displacement fields with only one or two effective quadrupoles, we apply radical Voronoi tessellation on each disk packing to obtain the dual Delaunay triangulations. For a packing with $N$ disks in a 2D periodic box, the Delaunay triangulation consists of $2N$ triangles with a total of $3N$ edges. The network formed by the disk centers and Delaunay edges provides a fully connected reference state for each disk packing. For an isostatic packing at low pressures, the number of interparticle contacts is $N_c=2N-1$, which matches the number of degrees of freedom, leaving $N_m=N+1$ missing contacts compared to the fully connected Delaunay network. As the pressure increases, the number of missing contacts decreases monotonically, as shown in Fig.~\ref{fig:probQuad} (b). At low pressures $p \lesssim 0.01$ near jamming onset, isolated effective quadrupoles are not observed. Once the fractional number of missing contacts $\langle N_m/(N+1)\rangle \lesssim 5\%$, the probability $P(R^2>0.7)$ increases rapidly. At the smallest values of $\langle N_m/(N+1)\rangle$, non-affine displacement fields with one or two effective quadrupoles become prevalent, with $P(R^2(1)>0.7)=33\%$ and $P(R^2(2)>0.7)=76\%$. 

These results raise an important question: What causes the formation of quadrupoles in the non-affine displacement fields of jammed disk packings? To address this question, in the next subsection, we analyze the special case in which only a single contact is missing in jammed disk packings relative to the fully connected Delaunay network and relate the occurrence of quadrupolar displacement fields in jammed disk packings to the Eshelby inclusion and inhomogeneity problems in continuum materials.  

\begin{figure}[t]
    \centering
    \includegraphics[width=0.9\linewidth]{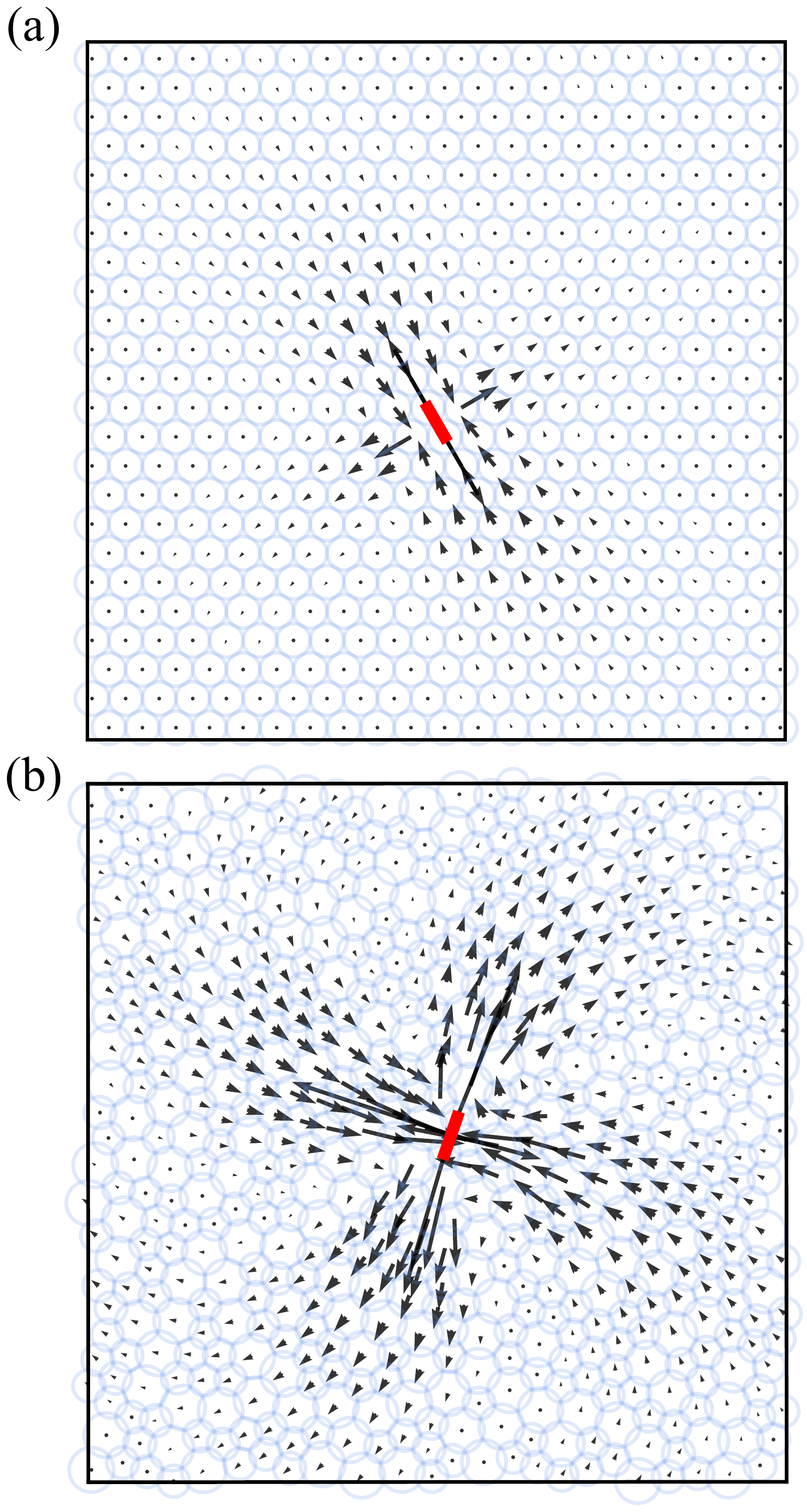}
    \caption{Non-affine displacement field ${\vec u}$ (black arrows) in response to a single simple shear step $\Delta \gamma=10^{-5}$ (followed by energy minimization) applied to jammed disk packings: (a) A nearly crystalline packing at pressure $p=1.8\times10^{-3}$, with a polydispersity of $\Delta \sigma/\langle \sigma\rangle = 6 \times 10^{-4}$ in diameter and a packing fraction of $\phi=0.9085$; and (b) A disordered bidisperse disk packing at $p=0.285$ and $\phi =1.2$. In both cases, the system contains $N_c = 3N-1$ interparticle contacts, i.e., one missing contact (highlighted by a red thick line) compared to the fully connected Delaunay network. The angle between the missing contact and the shear direction is (a) $\alpha = 2\pi/3$ and (b) $0.4\pi$.}
    \label{fig:xtalamph}
\end{figure}

\begin{figure}[t]
    \centering
    \includegraphics[width=\linewidth]{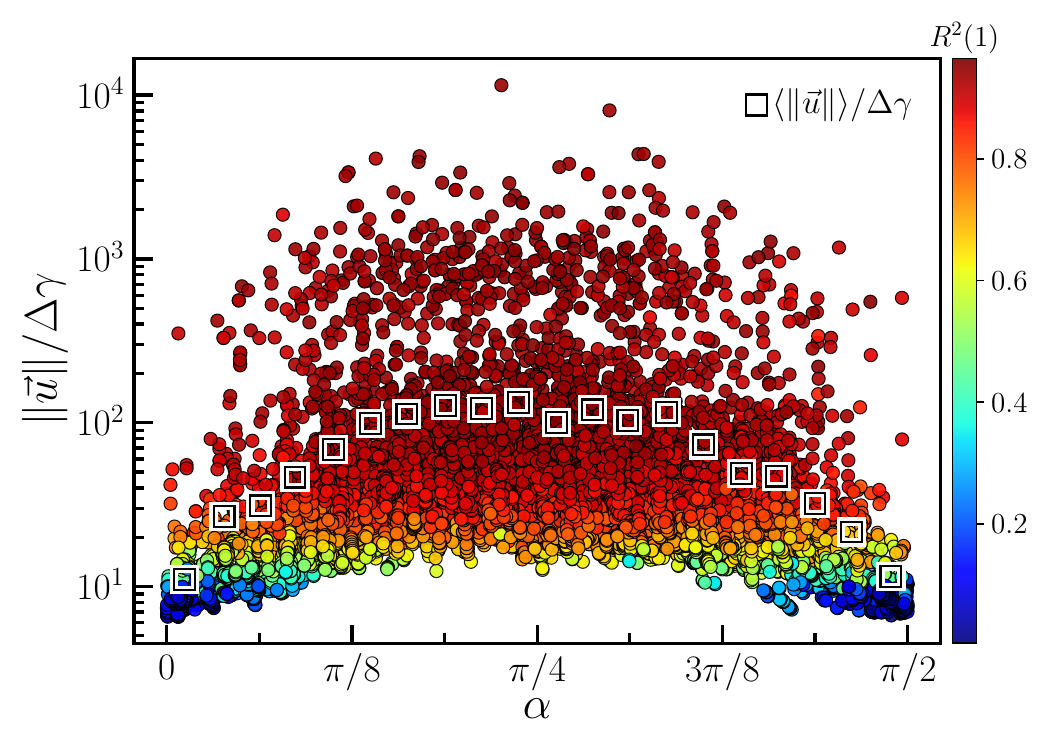}
    \caption{Magnitude of the non-affine displacement field $\|\vec{u}\|$, normalized by the simple shear strain increment $\Delta\gamma$, for jammed disk packings with one missing contact (relative to the Delaunay-triangulated network) plotted as a function of the angle $\alpha$ of the missing bond relative to the shear direction. The circles are colored by $R^2(1)$ (increasing from violet to dark red) from fits of ${\vec u}$ to the displacement field for a single effective quadrupole (Eq.~\ref{fiteq}). The squares indicate the average $\langle \|\vec{u}\|\rangle /\Delta \gamma$ at each $\alpha$.}
    \label{fig:mag_orient}
\end{figure}

\subsection{Eshelby inclusion and inhomogeneity problems}
\label{results:B}

In Sec.~\ref{results:A}, we showed that isolated quadrupolar displacement fields are observed across a range of pressures in jammed disk packings, but they are most frequently found in the limit of high pressures (cf. Fig.~\ref{fig:probQuad}). Also, prior studies have shown that the non-affine displacement fields in amorphous solids that arise from local rearrangements during applied shear often resemble the quadrupolar-like solutions to the continuum Eshelby inclusion problem~\cite{kang2023, maloney04, maloney06, jin21,sopu21,zhang21, dasgupta12, desmarchelier24}, where a uniform instantaneous plastic strain is applied to an elliptical inclusion within an infinite elastic material. Elliptical regions undergoing instantaneous plastic strains in continuum solids arise in applications ranging from thermal expansion and twinning in crystals~\cite{eshelby1957}, martensitic transformations~\cite{VASOYA2023}, and the mechanical response of concrete~\cite{GUO2024}. Solutions to the Eshelby inclusion problem can also be used to solve the Eshelby inhomogeneity problem, where the stiffness of the elliptical region differs from that of the surrounding matrix material. To further illustrate the analogy between the quadrupolar displacement fields observed in jammed disk packings and continuum solutions to the Eshelby inclusion and inhomogeneity problems, we examine two special cases of jammed disk packings, as shown in Fig.~\ref{fig:xtalamph}, that contain only one missing contact with respect to the fully connected network generated from Delaunay triangulation of the disk centers. 

In Fig~\ref{fig:xtalamph} (a), we show the displacement field after a single simple shear strain step (followed by energy minimization) applied to a nearly crystalline packing with a small polydispersity in disk diameters $\Delta \sigma/\langle \sigma\rangle = 6 \times 10^{-4}$ and one missing contact that is oriented at an angle $\alpha = 2\pi/3$ relative to the shear direction. The displacement field is quadrupolar, centered at the missing contact, and oriented along the direction of the missing contact. Similar to the Eshelby inhomogeneity problem for continuum materials, the single missing contact in the discrete system induces a stark mismatch in elastic properties between the site of the missing bond and the rest of the system. A similar phenomenon is observed in disordered jammed disk packings, as shown in Fig.~\ref{fig:xtalamph} (b). Despite the amorphous structure, a missing contact (relative to the fully connected Delaunay network) gives rise to a quadrupolar displacement field centered at the missing contact.

Since the stiffness tensor contains different components that correspond to different modes of deformation, in general, not all mismatches in local stiffness are activated by a given applied strain. For example, a quadrupolar displacement field will only be triggered when the region of the local stiffness mismatch is activated by the applied strain. In Fig.~\ref{fig:mag_orient}, we show the magnitudes of the non-affine displacement field $\|{\vec u}\|/\Delta \gamma$ in disk packings with a single missing contact (relative to the Delaunay triangulated network) after a single simple shear strain step plotted as a function of the orientation $\alpha$ of the missing bond. 
The average magnitude $\langle \| {\vec u}\|\rangle/\Delta \gamma$ has a peak at $\alpha = \pi/4$ and is minimal at $\alpha = 0$ and $\pi/2$.  
Large non-affine displacements do not occur when the missing bond is either parallel or perpendicular to the shear direction. As a result, $R^2(1) \sim 0$ for $\alpha = 0$ and $\pi/2$ when fitting $\vec{u}$ to the non-affine displacement field for a single effective quadrupole, indicating that quadrupoles are not activated at these angles. 
The emergence of quadrupolar displacement fields centered at missing contacts and aligned with the shear direction supports a theoretical description of non-affine displacement fields in jammed disk packings that is based on Eshelby's inclusion and inhomogeneity problems.

\begin{figure}[t]
    \centering
    \includegraphics[width=\linewidth]{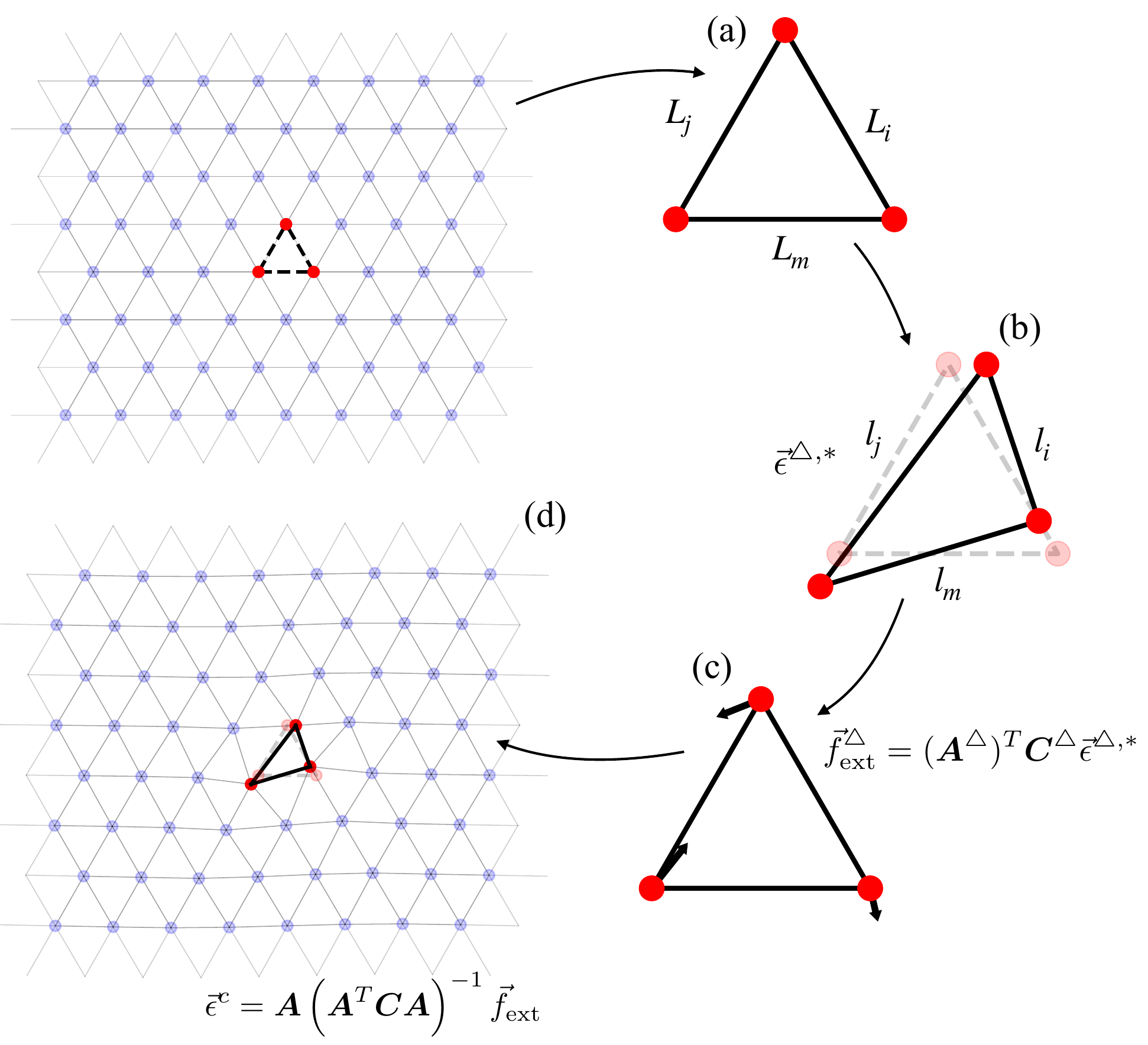}
    \caption{A schematic that describes the Eshelby inclusion problem for a triangular spring network. (a) A triangular inclusion is removed from the spring network with initial equilibrium rest lengths $L_{i0} = L_i$. (b) As an example, a pure shear eigenstrain $\vec{\epsilon}^{\Delta,*}$ is applied to the triangle and the equilibrium rest lengths of the springs are set to the current triangle edge lengths $L_{i0}=l_i$ (solid lines). The original triangle is indicated by the dashed lines. (c) An external force $\vec{f}_{\rm ext}^\Delta = (\boldsymbol{A}^{\triangle})^T\boldsymbol{C}^\triangle\vec{\epsilon}^{\triangle,*}$ (Eq.~\ref{eq:f_from_u}) is applied to the vertices to push the triangle back into its original shape, where $\boldsymbol{C}^\triangle$ and $\boldsymbol{A}^{\triangle}$ are the stiffness and gradient matrices for the triangle. (d) After reinserting the triangle back into the network and allowing the system to relax, the triangle will have a new strain $\vec{\epsilon}^{\Delta,c}$ (indicated by the solid black lines), which can be calculated from $\vec{\epsilon}^{c}=\boldsymbol{A}\left(\boldsymbol{A}^T\boldsymbol{C}\boldsymbol{A}\right)^{-1}\vec{f}_{\rm ext}$. The global strain vector $\vec{\epsilon}^{c}$ contains the strains of all triangles in the network (Eq.~\ref{eq:global_strain}). ${\vec f}_{\rm ext}$ is a global force $2N\times1$ vector containing forces $\vec{f}_{\rm ext}^\triangle$ on the red nodes in (d) and zeros elsewhere. $\boldsymbol{A}$ is the global gradient matrix defined in terms of $\boldsymbol{A}^{\triangle}$ in Eq.~\ref{eq:construct_A}.}
    \label{fig:discrete_inclusion}
\end{figure}

\begin{figure*}[t]
    \centering
    \includegraphics[width=\linewidth]{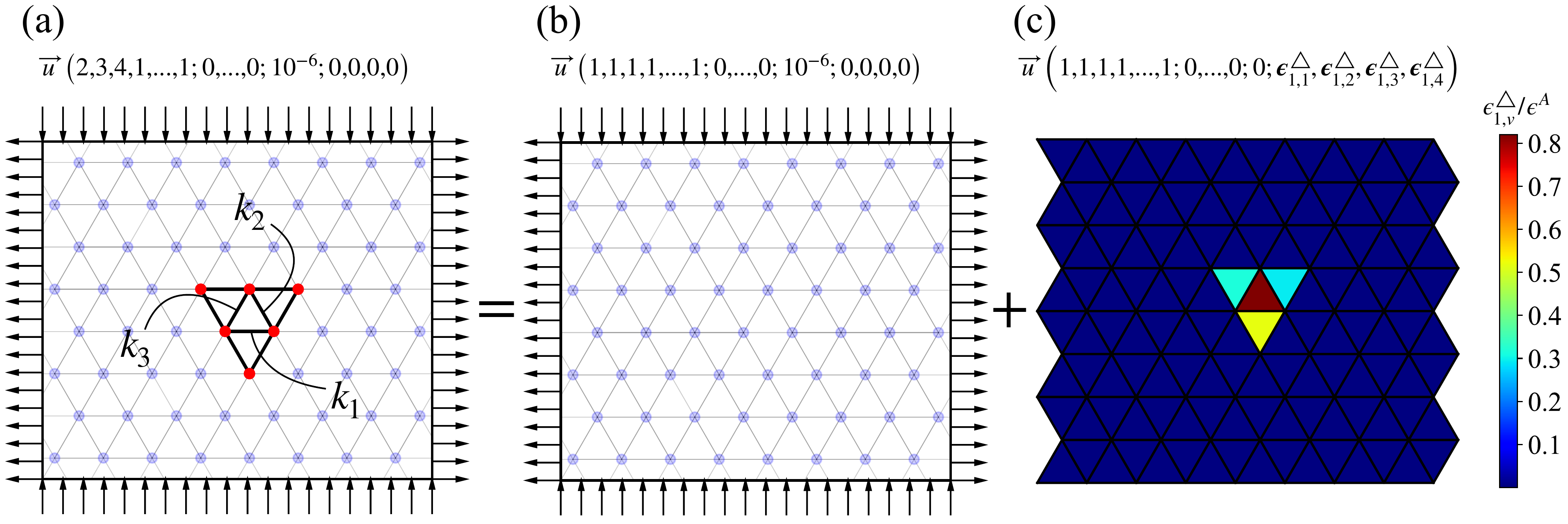}
    \caption{A schematic illustrating the displacement field solution, $\vec{u}\left(k_1/k_0,k_2/k_0,k_3/k_0,k_4/k_0,...,k_{3N}/k_0; t_1,...,t_{3N};  \epsilon^A;  \epsilon_{1,1}^\triangle,  \epsilon_{1,2}^\triangle,  \right.$ $\left. \epsilon_{1,3}^\triangle, \epsilon_{1,4}^\triangle \right)$, to the Eshelby inhomogeneity problem for a triangle with three modified spring constants in a spring network under a global pure shear. (a) We decompose the displacement field solution to a global pure shear strain $\boldsymbol{\epsilon}^A$ of magnitude $\epsilon^A =10^{-6}$ applied to a spring network containing three bonds with spring constants ($k_1/k_0=2$, $k_2/k_0=3$, and $k_3/k_0=4$) that differ from those of the rest of the network $k_0$, into (b) the displacement field from the global strain $\boldsymbol{\epsilon}^A$ on the spring network where all spring constants are $k_0$ (Eq.~\ref{eq:u_R}) plus (c) the displacement field resulting from local eigenstrains $\epsilon^\triangle_{1,1}$, $\epsilon^\triangle_{1,2}$, $\epsilon^\triangle_{1,3}$, and $\epsilon^\triangle_{1,4}$ on the four triangles that share the modified bonds, where again all spring constants are $k_0$, (Eq.~\ref{eq:u_star}). In (c), we show the von Mises strain of the local eigenstrain matrices (cf. Eq.~\ref{eq:vm}). In the case presented here, the spring network does not have pre-stress, and thus the bond tensions $t_1=0,...,t_{3N}=0$.}
    \label{fig:discrete_inhomogeneity}
\end{figure*}

The occurrence of quadrupoles in jammed disk packings raises an important question: Can the Eshelby inclusion problem for continuum materials be reformulated for discrete particulate systems to explain the emergence of isolated quadrupolar displacement fields, as well as more general disordered displacement fields? A key challenge in addressing this question is the fact that jammed disk packings exhibit locally varying elastic moduli~\cite{zhang23}, and thus multiple interacting Eshelby inclusions with differing stiffness must be embedded within the background material. To lay the groundwork for a complete solution to the discrete Eshelby inclusion and inhomogeneity problems, we begin by introducing the continuum Eshelby inclusion and inhomogeneity problems and then discuss how they can be reformulated in the case of particulate systems, such as jammed disk packings. 

The classical Eshelby inclusion problem assumes that an elliptical region within an infinite, elastic material is removed from the matrix material and undergoes a known uniform, plastic ``eigenstrain'' $\epsilon_{ij}^*$\cite{eshelby1957}, specified by three values for symmetric strains in two dimensions. A traction is then applied to the surface of the isolated, strained inclusion to deform it back into its original shape. The inclusion is then reinserted into the matrix and the system is allowed to relax. Due to the surrounding elastic matrix, stresses are generated in both the inclusion and matrix, yielding a final, or ``constrained'' strain $\epsilon_{ij}^{c}$ that differs from $\epsilon_{ij}^*$.  The constrained strain and plastic eigenstrain are related via the Eshelby tensor: $\epsilon_{ij}^{c} = S_{ijkl}\epsilon_{kl}^*$, where $S_{ijkl}$ is a function of the stiffness matrix of the material and geometrical properties of the elliptical inclusion \cite{MENG2012}. (See Fig.~\ref{fig:continuum_inclusion} in Appendix A for additional details.)

The Eshelby inclusion problem can also be formulated for jammed disk packings. For example, we can form a spring network based on the Delaunay triangulation of the centers of jammed disk packings.
The analog in a discrete particulate system to an elliptical inclusion in a continuum system is a Delaunay triangle of the spring network, since a triangle contains the minimum number of degrees of freedom to define a stress and strain tensor in two dimensions. A schematic describing similar operations to determine the constrained strain for the Eshelby inclusion problem in a triangular spring network is shown in Fig.~\ref{fig:discrete_inclusion}. We first remove a triangle from the network (panel (a)), plastically deform the triangle by the strain $\vec{\epsilon}^{\Delta^*}$, which is a row vector with four entries to describe non-symmetric strains locally (panel (b)), then apply forces to the vertices to push the triangle back into its original shape (panel (c)), reinsert it back into the network, and allow the system to relax (panel (d)). The final strain of the triangle, $\vec{\epsilon}^{\Delta^c}$, will be different from $\vec{\epsilon}^{\Delta^*}$ due to the surrounding network. 

The solution to the Eshelby inclusion problem for continuum materials can be used to construct the displacement field for the Eshelby inhomogeneity, or elastic mismatch, problem. The Eshelby inhomogeneity problem embeds an ellipsoidal inclusion within an infinite elastic matrix that has a different stiffness tensor from that of the inclusion, and an affine strain is imposed on the system. (See Fig.~\ref{fig:continuum_inhomogeneity} in Appendix A.) Using Eshelby's equivalent inclusion method (EIM), the displacement field solution to the Eshelby inhomogeneity problem can be written as a sum of the displacement field from the affine strain $\epsilon_{kl}^A$ on the elastic matrix without the inclusion plus the displacement field solution to the Eshelby inclusion problem with eigenstrain determined by
\begin{align}
\label{eq:EIM}
\left[\left(C^{I}_{ijkl} - C^{M}_{ijkl}\right)S_{klmn} + C^{M}_{ijmn}\right]\epsilon^*_{mn} = \nonumber \\ \left(C^{M}_{ijkl} - C^{I}_{ijkl}\right)\epsilon^A_{kl},
\end{align}
where $C^{M}_{ijkl}$ and $C^{I}_{ijkl}$ are the stiffness tensors for the matrix and inclusion, respectively \cite{mura2013micromechanics}. Note that the eigenstrain is applied to the reference system with uniform stiffness tensors. 

We can also apply a similar methodology for determining the displacement field for a discrete system (i.e. a spring network) under an affine pure shear $\boldsymbol{\epsilon}^A$, where the spring constants of the three bonds in a single triangle differ from those of the rest of the network. 
A schematic showing the EIM solution procedure to the Eshelby inhomogeneity problem in a triangular spring network is shown in Fig.~\ref{fig:discrete_inhomogeneity}. We first introduce an elastic mismatch by resetting the spring constants ($k_1/k_0=2$, $k_2/k_0=3$, and $k_3/k_0=4$) of three springs (panel (a)), which changes the stiffness matrices of the four triangles that share the bonds with modified spring constants. Note that after introducing this elastic mismatch, the network is still in equilibrium. An affine pure shear strain $\boldsymbol{\epsilon}^A$ is then applied and the elastic mismatch will generate a non-affine displacement field $\vec{u}$. The total displacement field is the sum of the global strain of the uniform spring network (panel b) (affine contribution) plus four eigenstrain perturbations applied to the triangles where all bonds are uniform with spring constant $k_0=k$ (non-affine contribution).

\subsection{Eshelby's equivalent inclusion method for jammed disk packings}
\label{results:C}

In Sec.~\ref{results:B}, we showed that Eshelby's equivalent inclusion method (EIM) can be used to solve for the non-affine displacement field in continuum materials that contain a single inclusion (with a different stiffness tensor from the background matrix) and are under an applied affine strain. In particular, applying the eigenstrain $\epsilon^*_{mn}$ in Eq.~\ref{eq:EIM} to the inclusion (where all elastic properties are uniform) yields the correct displacement field.  Similar to the Eshelby inclusion and inhomogeneity problems in continuum materials, we can consider a spring network where the spring constants of a single triangle have been changed relative to the rest of the network, as shown in Fig.~\ref{fig:discrete_inhomogeneity}.  This change in stiffness gives rise to local eigenstrains on the central and neighboring triangles. In this subsection, we present a re-formulation of the EIM that enables the calculation of the eigenstrains that arise from differences in the triangle stiffnesses.
Note that the continuum EIM is only formulated for systems containing an isolated inclusion. The EIM is generally not used for multiple, interacting inclusions in continuum materials~\cite{sarkar2021elastic,HORII1985}.
However, the reformulated EIM described below is valid for an arbitrary number of interacting triangle inclusions in particulate systems, such as jammed disk packings. 

In the remainder of this subsection, we introduce a methodology that will allow us to convert between single triangle strains and stresses and between node forces and displacements. To do this, we will calculate the individual triangle stiffness matrices for the original jammed packing and its reference network. We then derive the global stiffness matrix (relating stiffness to force) and gradient matrix (relating displacement to strain) for the jammed packing and its reference network. We use these global quantities to develop an EIM-like approach for jammed disk packings that allows us to calculate the eigenstrains, which reconstruct the non-affine displacement field of the jammed disk packing for a given applied affine strain.

\subsubsection{Calculation of the stiffness matrices for triangles}

\begin{figure}[h]
    \centering
    \includegraphics[width=0.9\linewidth]{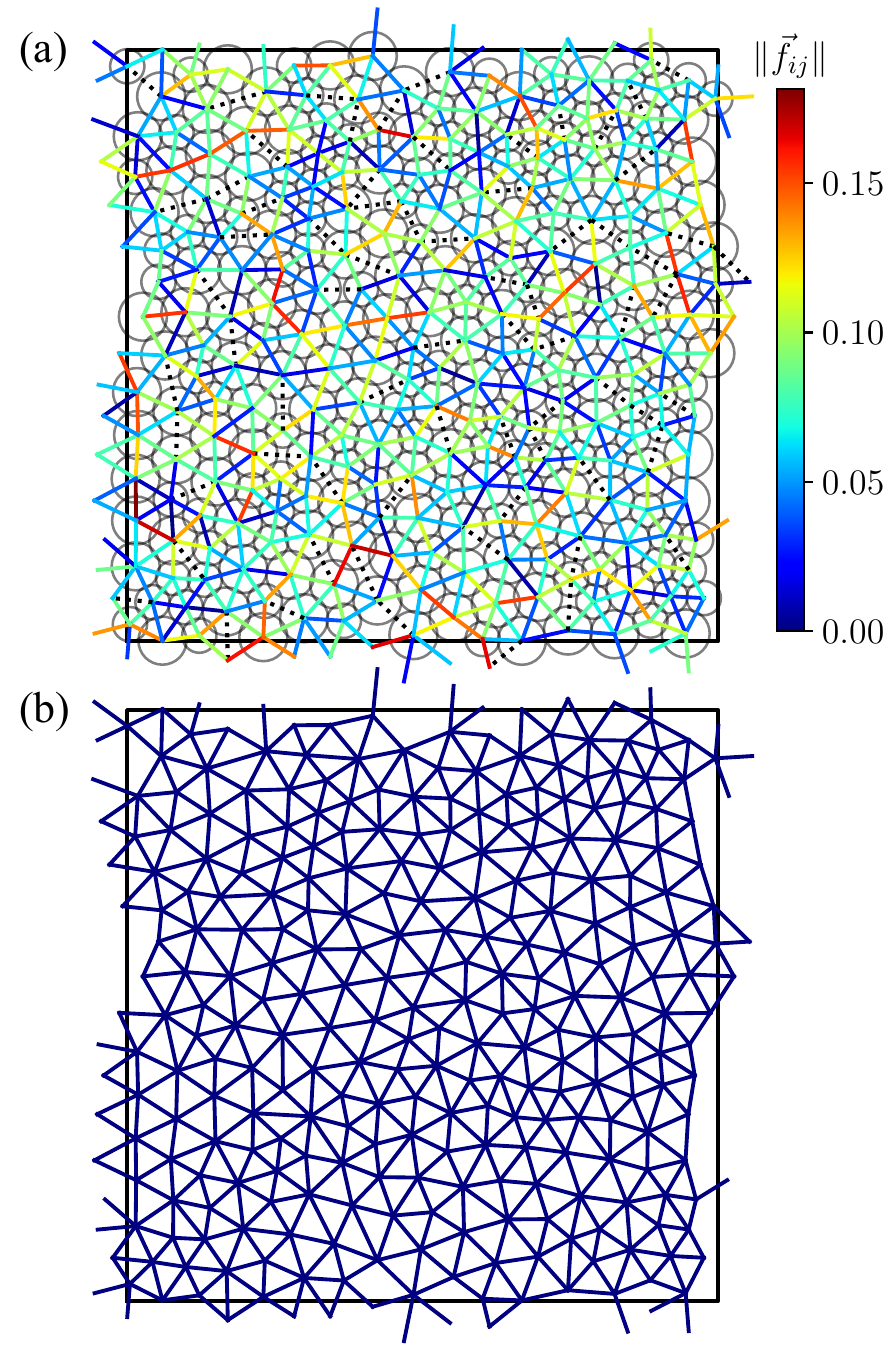}
    \caption{(a) A jammed bidisperse disk packing with $N=256$ at pressure $p=0.1$, $\gamma=0$, and fraction of missing contacts $N_m/(N+1)\approx 0.27$. The interparticle contacts are color-coded by $||\vec{f}_{ij}||$. The missing contacts are highlighted by dotted lines. (b) The reference spring network obtained from the Delaunay triangulation of the disk packing in (a).  The equilibrium bond lengths in (b) are set to the interparticle separations in the network such that $\|\vec{f}_{ij}\|\equiv0$.}
    \label{fig:ref}
\end{figure}

To develop the EIM-like approach for jammed disk packings, we first need to generate a reference ``elastic matrix'' for each jammed packing at a given pressure $p$ and shear strain $\gamma$. To do this, we perform a Delaunay triangulation of the disk centers at each $p$ and $\gamma$. We form a spring network with uniform spring constants $k_0=k$ using the connectivity of the triangulated network with equilibrium bond lengths set to the bond lengths of the network. Thus, for each jammed packing, we have the original configuration and its reference network as shown in Fig.~\ref{fig:ref} (a) and (b). 

\begin{figure}[h]
\centering
\includegraphics[width=0.3\textwidth]{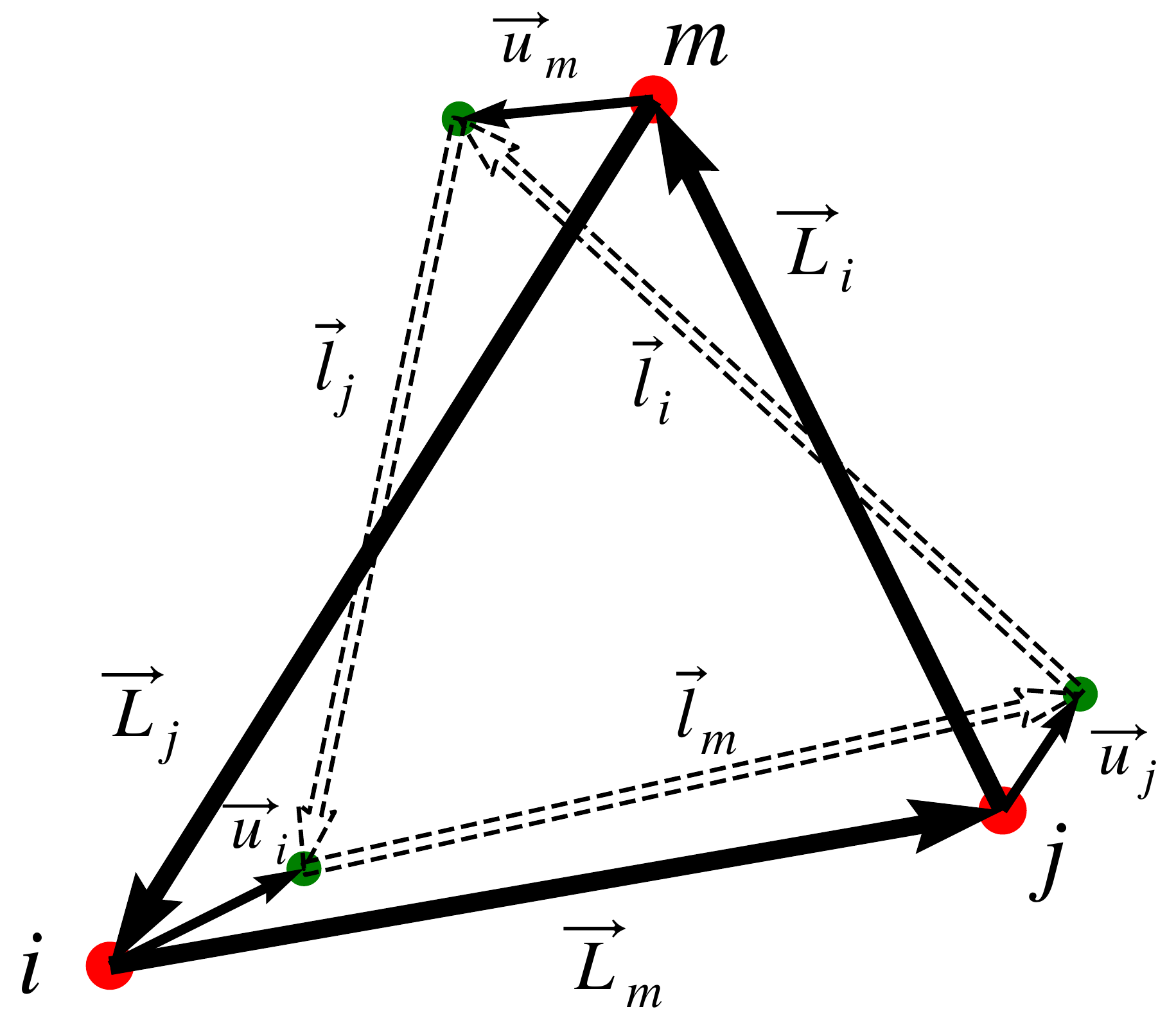}
\caption{Undeformed triangle with sides (solid arrows) defined by the vectors $\vec{L}_i$, $\vec{L}_j$, and $\vec{L}_m$. After the nodes $i$, $j$, and $m$ are are displaced by $\vec{u}_i$, $\vec{u}_j$, and $\vec{u}_m$, the deformed sides (dashed arrows) are defined by $\vec{l}_i$, $\vec{l}_j$, and $\vec{l}_m$.}
\label{triangle_fig}
\end{figure}

To calculate the triangle stiffness matrix, we will first define the strain for a triangle with nodes $i$, $j$, and $m$ and sides defined by the vectors ${\vec L}_i$, ${\vec L}_j$, and ${\vec L}_m$, as shown in Fig.~\ref{triangle_fig}. We will then express the potential energy of a triangle $U^{\triangle}$ in terms of the bond elongations and then the triangle strain, expand the potential energy in terms of the triangle strain, and identify the coefficients of the second-order terms in strain as the stiffness matrix of the triangle.

Using Eq.~\ref{potential}, we can write the potential energy for a triangle in terms of interactions between each pair of vertices (or nodes),
\begin{equation}
\label{potential2}
    U^\triangle = U_{mj}(r_{mj}) + U_{im}(r_{im}) + U_{ji}(r_{ji}).
\end{equation}
We can then rewrite $U^\triangle$ in terms of the bond elongations $e_i$ that are generated by the displacements of the triangle nodes:  
\begin{equation}
\label{potential3}
    U^\triangle = \frac{1}{2}k_ie_i^2 + \frac{1}{2}k_je_j^2 + \frac{1}{2}k_me_m^2,
\end{equation}
where the bond elongations are 
\begin{equation}
    \label{eq:elong0}
    e_i = \left\lvert \vec{r}_{mj} -\vec{u}_j+\vec{u}_m\right\lvert  - \sigma_{mj},
\end{equation}
and the spring constants are divided by two, $k_i=k_{mj}/2$, $k_j=k_{im}/2$, $k_m=k_{ji}/2$, since each bond is shared by two adjacent triangles. In  Eq.~\ref{potential3}, we assume that since the jammed disk packings are overcompressed, they experience an effective linear spring interaction potential with both attractive and repulsive interactions. Defining the vectors $\vec{l}_i=\vec{r}_{mj}-\vec{u}_j+\vec{u}_m$ and $\vec{L}_i=\vec{r}_{mj}$ to represent the deformed and undeformed node positions of side $i$, respectively, we can rewrite the elongations as 
\begin{equation}
    \label{eq:elong}
    e_i = \left\lvert \vec{l}_i \right\lvert  - \left\lvert \vec{L}_i \right\lvert + b_i,
\end{equation}
where
\begin{equation}
    b_i = L_i - L_{i_0}
\end{equation}
and $L_{i_0}$ is the equilibrium length of the spring associated with side $i$. 
However, we now need to relate the bond elongations to the triangle strain: 
\begin{equation}
    \label{eq:strain}
    \boldsymbol{\epsilon}^\triangle =\begin{bmatrix} 
\epsilon_{xx} &\epsilon_{xy} \\
\epsilon_{yx} &\epsilon_{yy} 
\end{bmatrix}
= \boldsymbol{F}^\triangle - \boldsymbol{I}, 
\end{equation}
where the $2 \times 2$ deformation gradient matrix for the triangle $\boldsymbol{F}^\triangle$ can be expressed using any two sides of the deformed and undeformed triangles,
\begin{equation}
    \label{eq:def_grad}
    \boldsymbol{F}^{\triangle} = \begin{bmatrix}
    \vec{l}_i & \vec{l}_j
    \end{bmatrix}
    \begin{bmatrix}
    \vec{L}_i & \vec{L}_j
    \end{bmatrix}^{-1}
\end{equation}
and $\boldsymbol{I}$ is the $2\times 2$ identity matrix. 

We can express the elongations in terms of the triangle deformation gradient $\boldsymbol{F}^\Delta$ by calculating the change in the square of the length of side $i$ of the triangle:
\begin{subequations}
\begin{align}
   \left\lvert \vec{l}_i \right\lvert^2  - \left\lvert \vec{L}_i \right\lvert^2 &= \left(\boldsymbol{F}^{\triangle}\vec{L}_i\right)\cdot\left(\boldsymbol{F}^{\triangle}\vec{L}_i\right) - \vec{L}_i\cdot\vec{L}_i \\
   &= \vec{L}_i^T\left((\boldsymbol{F}^{\triangle})^T\boldsymbol{F}^\Delta-\boldsymbol{I} \right)\vec{L}_i.
\label{eq:change_sq}
\end{align}
\end{subequations}
By rearranging Eq.~\ref{eq:elong}, we can substitute  
\begin{equation}
    \left\lvert \vec{l}_i \right\lvert  = e_i +  \left\lvert \vec{L}_i \right\lvert - b_i
\end{equation}into Eq.~\ref{eq:change_sq}, which yields
\begin{equation}
\label{step}
\left(e_i+\left\lvert\vec{L}_i\right\lvert-b_i\right)^2  = \vec{L}_i^T\left((\boldsymbol{F}^{\triangle})^T\boldsymbol{F}^{\triangle}-\boldsymbol{I} \right)\vec{L}_i + \left\lvert \vec{L}_i \right\lvert^2.
\end{equation}
After solving Eq.~\ref{step} for $e_i$, we find 
\begin{equation}
e_i = \left(\sqrt{1 + \hat{L}_i^T\left((\boldsymbol{F}^{\triangle})^T\boldsymbol{F}^{\triangle}-\boldsymbol{I} \right)\hat{L}_i } - 1\right)L_i + b_i
\end{equation}
The potential energy in Eq.~\ref{potential3} for the triangle with nodes $i$, $j$, and $m$ can now be rewritten in terms of $\boldsymbol{F}^{\triangle}$:
\begin{widetext}
\begin{equation}
    \label{eq:U_wrt_F}
    U^{\triangle} = \sum_{\alpha} \frac{1}{2}k_{\alpha}e_{\alpha}^2
    = \sum_{\alpha} \frac{1}{2}k_{\alpha}\left[\left(\sqrt{1+2\hat{L}_{\alpha}^T\left[ \frac{1}{2}\left((\boldsymbol{F}^{\triangle})^T\boldsymbol{F}^{\triangle}-\boldsymbol{I}\right)\right]\hat{L}_{\alpha}} - 1\right)L_\alpha+b_{\alpha}\right ]^2,
\end{equation}
where $\alpha$ sums over the nodes $i$, $j$, and $m$.
\end{widetext}
Expanding the potential energy of the deformed triangle to second order in $(\boldsymbol{F}^{\triangle} - \boldsymbol{I})$, we obtain
\begin{eqnarray}
    \label{eq:expansion}
    U^{\triangle}\left(\boldsymbol{F}^{\triangle}-\boldsymbol{I}\right) & =  U^{\triangle}(0) + \frac{\partial U^{\triangle}}{\partial\boldsymbol{F}^{\triangle}} \Bigg\rvert_{\boldsymbol{F}^{\triangle} = \boldsymbol{I}}(\boldsymbol{F}^{\triangle}-\boldsymbol{I}) + \\
   &  \frac{1}{2}(\boldsymbol{F}^{\triangle}-\boldsymbol{I})^T\frac{\partial^2 U^{\triangle}}{\partial(\boldsymbol{F}^{\triangle})^T\partial\boldsymbol{F}^{\triangle}}\Bigg\rvert_{\boldsymbol{F}^{\triangle} = \boldsymbol{I}} (\boldsymbol{F}^{\triangle}-\boldsymbol{I}). \nonumber
\end{eqnarray}
The triangle stiffness tensor
\begin{equation}
    \label{eq:stiffness_tensor}
    \boldsymbol{C}^\triangle = \frac{\partial^2 U^{\triangle}}{\partial(\boldsymbol{F}^{\triangle})^T\partial\boldsymbol{F}^{\triangle}}\Bigg\rvert_{\boldsymbol{F}^{\triangle} = \boldsymbol{I}}
\end{equation}
is a $2\times2\times2\times2$ tensor that can be written as a $4\times4$ matrix,
\begin{equation}
    \label{eq:triangle_C_mat}
\boldsymbol{C}^\triangle = \begin{bmatrix}
        c_{xxxx} & c_{xxyy} & c_{xxxy} & c_{xxyx} \\
        c_{yyxx} & c_{yyyy} & c_{yyxy} & c_{yyyx} \\
        c_{xyxx} & c_{xyyy} & c_{xyxy} & c_{xyyx} \\
        c_{yxxx} & c_{yxyy} & c_{yxxy} & c_{yxyx}
    \end{bmatrix},
\end{equation}
where the components are given in Appendix~\ref{appendix:B}. 

We will now introduce the methodology for expressing the nodal displacements $\vec{u}_i$, $\vec{u}_j$, and $\vec{u}_m$ (Fig.~\ref{triangle_fig}) in terms of the triangle strain $\boldsymbol{\epsilon}^\triangle$ (Eq.~\ref{eq:strain}) using the gradient matrix $\boldsymbol{A}^\triangle$ and for relating $\boldsymbol{\epsilon}^\triangle$ to the triangle stress $\boldsymbol{P}^\triangle$ via the stiffness tensor (Eq.~\ref{eq:stiffness_tensor}). Finally, we show that the forces on the triangle nodes can be expressed in terms of $(\boldsymbol{A}^\triangle)^T$ and $\boldsymbol{P}^\triangle$. These expressions will be important for determining the triangle eigenstrains that when applied to the reference network generate the nonaffine displacement field. 

To obtain the relation between the triangle strain and nodal displacements, we substitute $\boldsymbol{F}^{\triangle}$ (Eq.~\ref{eq:def_grad}) into the definition of the triangle strain
in Eq.~\ref{eq:strain}:
\begin{equation}
    \label{eq:strain_l_L}
    \boldsymbol{\epsilon}^\triangle =  \begin{bmatrix}
    \vec{l}_i & \vec{l}_j
    \end{bmatrix}
    \begin{bmatrix}
    \vec{L}_i & \vec{L}_j
    \end{bmatrix}^{-1} - \boldsymbol{I}.
\end{equation}
The deformed sides $\vec{l}_i$, $\vec{l}_j$, and $\vec{l}_m$, can be expressed in terms of the nodal displacements using
\begin{equation}
    \label{eq:l_L_u}
    \vec{l}_i = \vec{L}_i + \boldsymbol{\nabla}
    \begin{bmatrix}
        \vec{u}_j \\ 
        \vec{u}_m
    \end{bmatrix},
\end{equation}
where 
\begin{equation}
    \boldsymbol{\nabla} = \begin{bmatrix}
        -\boldsymbol{I} & \boldsymbol{I} 
    \end{bmatrix} = \begin{bmatrix}
    -1 & 0 & 1 & 0 \\
    0  & -1 & 0 & 1
    \end{bmatrix}.
\end{equation}
We can then substitute Eq.~\ref{eq:l_L_u} into Eq.~\ref{eq:strain_l_L} to obtain
\begin{equation}
    \label{eq:strain_L_u}
    \boldsymbol{\epsilon}^\triangle = \begin{bmatrix}
    \vec{L}_i + \boldsymbol{\nabla}
    \begin{bmatrix}
        \vec{u}_j \\ 
        \vec{u}_m
    \end{bmatrix}
    &
    \vec{L}_j + \boldsymbol{\nabla}
    \begin{bmatrix}
        \vec{u}_m \\ 
        \vec{u}_i
    \end{bmatrix}
    \end{bmatrix}
    \begin{bmatrix}
    \vec{L}_i & \vec{L}_j
    \end{bmatrix}^{-1} - \boldsymbol{I}.
\end{equation}
Eq.~\ref{eq:strain_L_u} shows that the nodal displacements are linearly related to the triangle strain~\cite{madenci2015finite}:
\begin{equation}
    \label{eq:strain_disp}
    \vec{\epsilon}^\triangle = \boldsymbol{A}^\triangle\vec{u}^\triangle,
\end{equation}
where
\begin{equation}
    \label{eq:triangle_A}
    \boldsymbol{A}^\triangle = \frac{1}{2\mathcal{A}}\begin{bmatrix}
   -L_{yi} & 0 & L_{yj} & 0  & -L_{ym} & 0\\
   0 & L_{xi} & 0 & -L_{xj} & 0 & L_{xm} \\
   L_{xi} & 0 & -L_{xj} & 0 & L_{xm} & 0 \\
   0 & -L_{yi} & 0 & L_{yj} & 0 & -L_{ym}
 \end{bmatrix}
\end{equation}
is the $4 \times 6$ gradient matrix and 
\begin{equation}
    \mathcal{A} = \frac{1}{2}\left\lvert \vec{L}_i\times\vec{L}_j \right\lvert
\end{equation}
is the area of the undeformed triangle. In Eq.~\ref{eq:strain_disp}, we expressed the nodal displacements of the triangle as a $6\times1$ column vector
\begin{equation}
    \vec{u}^\triangle = \begin{bmatrix}
        \vec{u}_i^T & \vec{u}_j^T & \vec{u}_m^T
    \end{bmatrix}^T
\end{equation}
and used Voigt notation for the strain tensor, expressing it as a $4 \times 1$ column vector:
\begin{equation}
    \label{eq:mat_to_row}
    \vec{\epsilon}^{\triangle} = \begin{bmatrix}
    \epsilon_{xx} & \epsilon_{yy} & \epsilon_{xy} & \epsilon_{yx}
    \end{bmatrix}^T.
\end{equation}
Thus, $\boldsymbol{A}^\triangle$ acts as a gradient operator, converting displacements of single nodes to the strain of the {\it triangle}. The area term in the denominator of Eq.~\ref{eq:triangle_A} arises due to the inversion of $\begin{bmatrix} \vec{L}_i & \vec{L}_j
\end{bmatrix}$ in Eq.~\ref{eq:strain_L_u}. We adopt the convention that node entries in $\boldsymbol{A}^\triangle$ are ordered in a counter-clockwise fashion, i.e., columns $1$ and $2$ include node $i$, columns $3$ and $4$ include node $j$, and columns $5$ and $6$ include node $m$. 
 
The first Piola-Kirchhoff stress tensor for a triangle is obtained from the first derivative of the potential energy with respect to the deformation gradient matrix \cite{gil2016nonlinear},
\begin{equation}
    \boldsymbol{P}^\triangle =\frac{1}{\mathcal{A}}\frac{\partial U^{\triangle}}{\partial \boldsymbol{F}^\triangle},
\end{equation}
which is related to the Cauchy (virial) stress tensor~\cite{gil2016nonlinear}:
\begin{equation}
    \boldsymbol{\sigma}^\triangle = [\det(\boldsymbol{F}^\triangle)]^{-1}\boldsymbol{P}^\triangle(\boldsymbol{F}^{\triangle})^T.
\end{equation}
The first-order approximation to the first Piola-Kirchhoff stress resulting from the triangle strain $\vec{\epsilon}^\triangle$ can be obtained from the second and third terms of the expansion of the potential energy in $\boldsymbol{F}^{\triangle}$ (Eq.~\ref{eq:expansion}):
\begin{equation}
    \boldsymbol{{\hat P}}^\triangle_1 = \frac{1}{\mathcal{A}}\frac{\partial U^{\triangle}}{\partial \boldsymbol{F}^\Delta}\Bigg\rvert_{\boldsymbol{F}^{\triangle} = \boldsymbol{I}}  +\frac{1}{\mathcal{A}}\boldsymbol{C}^\triangle\vec{\epsilon}^\triangle,
\end{equation}
where the first term, originally constructed as a $2\times2$ matrix, can be converted into a $4\times1$ column vector using the convention
\begin{equation}
    \label{eq:mat_to_vec}
    \begin{bmatrix}
        a & b \\
        c & d
    \end{bmatrix} = 
    \begin{bmatrix}
        a & d & b & c
    \end{bmatrix}^T.
\end{equation}
We define
\begin{equation}
    \label{eq:delta_p}
    \Delta\boldsymbol{P}^\triangle \equiv \boldsymbol{{\hat P}}^\triangle_1 - \frac{1}{\mathcal{A}}\frac{\partial U^{\triangle}}{\partial \boldsymbol{F}^\triangle}\Bigg\rvert_{\boldsymbol{F}^{\triangle} = \boldsymbol{I}} =\frac{1}{\mathcal{A}}\boldsymbol{C}^\triangle\vec{\epsilon}^\triangle
\end{equation}
and relabel $\Delta\boldsymbol{P}^\triangle = \boldsymbol{P}_1^{\triangle}$ since the second term in Eq.~\ref{eq:delta_p} does not depend on ${\vec \epsilon}^{\triangle}$.  
We can then rewrite the change in energy of a triangle to first order in strain using Eq.~\ref{eq:delta_p}:
\begin{equation}
\label{new_deltaV}
    \Delta U^{\triangle} = U^{\triangle}\left(\boldsymbol{F}^{\triangle}-\boldsymbol{I}\right) - U^{\triangle}(0) = (\boldsymbol{F}^{\triangle}-\boldsymbol{I})^T \boldsymbol{P}_1^\triangle \mathcal{A}.
\end{equation}
Using Eqs.~\ref{eq:delta_p} and~\ref{new_deltaV}, the resulting nodal forces to first order in strain can be written as  
\begin{equation}
    \label{eq:force_eq}
    \vec{f}^\triangle = -\frac{\partial (\Delta U^{\triangle})}{\partial\vec{u}^\triangle} =  -\frac{\partial (\boldsymbol{F}^{\triangle})^T}{\partial\vec{u}^\triangle} \boldsymbol{P}_1^\triangle \mathcal{A} =-(\boldsymbol{A}^{\Delta})^T\vec{P}_1^\triangle\mathcal{A},
\end{equation}
where the Voigt notation is used to express the change in the first Piola-Kirchhoff stress
\begin{equation}
    \vec{P}_1^\triangle = \begin{bmatrix}
    P_{1xx}^{\triangle} & P_{1yy}^{\triangle} & P_{1xy}^{\triangle} & P_{1yx}^{\triangle} 
    \end{bmatrix}^T.
\end{equation}
Using Eqs.~\ref{eq:strain_disp} and~\ref{eq:delta_p}, we can rewrite the nodal forces in terms of the displacements
\begin{equation}
    \label{eq:f_from_u}
    \vec{f}^{\triangle} = -(\boldsymbol{A}^{\triangle})^T\boldsymbol{C}^\triangle\vec{\epsilon}^{\triangle} = -(\boldsymbol{A}^{\triangle})^T\boldsymbol{C}^\triangle\boldsymbol{A}^{\triangle}\vec{u}^\triangle,
\end{equation}
where $(\boldsymbol{A}^{\triangle})^T$ acts as the divergence operator in continuum mechanics. An external force $\vec{f}_{\rm ext}^\triangle=-\vec{f}^\triangle$ needs to be applied on the triangle to balance the force $\vec{f}^\triangle$ induced by the displacements $\vec{u}^\triangle$. Consequently, we can solve for the displacements using Eq.~\ref{eq:f_from_u}, 
\begin{equation}
    \label{eq:u_from_f}
    \vec{u}^\triangle = \left[(\boldsymbol{A}^{\triangle})^T\boldsymbol{C}^\triangle\boldsymbol{A}^{\triangle}\right]^{-1}\vec{f}_{\rm ext}^{\triangle}.
\end{equation}

We show how to construct the global gradient matrix $\boldsymbol{A}$, stiffness matrix $\boldsymbol{C}$, first Piola-Kirchhoff stress vector $\vec{P}_1$, strain vector $\vec{\epsilon}$, and define an area matrix $\boldsymbol{S}$ for the entire jammed packing from the corresponding local triangle properties in Appendix~\ref{appendix:C}. These global quantities are then used to formulate the EIM for spring networks in the next section.

\subsubsection{Equivalent inclusion method}
\label{eigen}

As described in Fig.~\ref{fig:discrete_inhomogeneity}, we will decompose the response to an affine shear strain applied to a jammed disk packing into the sum of two contributions ($\vec{u} = \vec{u}^R + \vec{u}^*$): 1) the displacement field ${\vec u}^R$ (associated with strain $\vec{\epsilon}^R$) arising from an affine simple shear strain $\vec{\epsilon}^A$  applied to the reference network of the jammed disk packing, and 2) the displacement field ${\vec u}^*$ arising from the eigenstrains $\vec{\epsilon}^{\triangle}_1$ applied to individual triangles in the reference network.

The strain $\vec{\epsilon}^R$ has two parts: the affine strain $\vec{\epsilon}^A$ applied to the reference network and the strain generated from $\vec{u}^R$ after relaxation of the reference network under the affine strain, i.e. 
\begin{equation}
\label{eq:eps_R}
 \vec{\epsilon}^R
= 
\boldsymbol{A}\vec{u}^R 
+
\vec{\epsilon}^{A},
 \end{equation}
where
\begin{equation}
    \vec{\epsilon}^A = \begin{bmatrix}
        (\vec{\epsilon}_{1}^{\triangle,A})^T & (\vec{\epsilon}_{2}^{\triangle,A})^T  & \cdots & (\vec{\epsilon}_{2N}^{\triangle,A})^T
    \end{bmatrix}^T
\end{equation}
and each triangle $\vec{\epsilon}^{\triangle,A}$ contains the entries of $\boldsymbol{\epsilon}^A$ (Eq.~\ref{eq:strain}) arranged as a vector according to Eq.~\ref{eq:mat_to_vec}. 

The displacement field $\vec{u}^R$, arising from the affine strain $\vec{\epsilon}^A$ applied to the reference network, can be calculated by finding the first Piola-Kirchhoff stress generated from ${\vec \epsilon}^A$ using Eq.~\ref{eq:delta_p}, and then finding the resulting forces from the stress using Eq.~\ref{eq:force_eq}.  Since the jammed packings are in force balance,
\begin{equation}
\label{temp}
    -\boldsymbol{A}^T\boldsymbol{C}_0\left(\boldsymbol{A}\vec{u}^R+\vec{\epsilon}^A\right) = \boldsymbol{0}.
\end{equation}
We can then solve for the displacements in Eq.~\ref{temp}, 
\begin{equation}
    \label{eq:u_R}
    \vec{u}^R = -\left(\boldsymbol{A}^T\boldsymbol{C}^0\boldsymbol{A}\right)^{-1}\boldsymbol{A}^T\boldsymbol{C}^0\vec{\epsilon}^{A}.
\end{equation}

For the reference network and jammed disk packing to possess the same mechanical response, we enforce that the total strain in the reference network $\vec{\epsilon}_{r}$ and in the jammed disk packing $\vec{\epsilon}_{p}$ are equal,  $\vec{\epsilon}_{r} = \vec{\epsilon}_{p} \equiv \vec{\epsilon}$. We also enforce that the stress in the reference network $\vec{P}_1^0$ and the jammed disk packing $\vec{P}_1^\epsilon$ are equal. 
The stress in the reference network is given by the stress associated with the total strain $\vec{P}_{1}^{0,\epsilon}$ minus the stress associated with the eigenstrain $\vec{P}_{1}^{0,*}$:
\begin{equation}
\label{eq:p_ref}
    \vec{P}_1^0 = \vec{P}_{1}^{0,\epsilon} - \vec{P}_{1}^{0,*} = \boldsymbol{C}^0\vec{\epsilon}\boldsymbol{S}^{-1}-\vec{P}_{1}^{0,*}.
\end{equation}
The stress in the jammed disk packing is generated from elastic strains, 
\begin{equation}
\label{eq:p_pack}
     \vec{P}_{1}^\epsilon = \boldsymbol{C}\vec{\epsilon}\boldsymbol{S}^{-1}.
\end{equation}
Equating the stresses in the reference network and the jammed disk packing yields
\begin{equation}
\label{minussign}
    \vec{P}_{1}^{0,\epsilon} - \vec{P}_{1}^{0,*} = \vec{P}_{1}^\epsilon.
\end{equation}
We then solve Eq.~\ref{minussign} for the stresses associated with the eigenstrain 
\begin{equation}
    \vec{P}_{1}^{0,*} = \left(\boldsymbol{C}^0 - \boldsymbol{C}\right)\vec{\epsilon}\boldsymbol{S}^{-1},
\end{equation}
where the total strain ${\vec \epsilon}$ can be obtained by applying the affine strain ${\vec \epsilon^A}$ to the jammed disk packing and allowing it to relax.  Analogous to Eqs.~\ref{eq:eps_R} and~\ref{eq:u_R}, after relaxation, the total strain in the jammed disk packing is
\begin{equation}
    \vec{\epsilon} =-\boldsymbol{A}\left(\boldsymbol{A}^T\boldsymbol{C}\boldsymbol{A}\right)^{-1}\boldsymbol{A}^T\boldsymbol{C}\vec{\epsilon}^{A} +\vec{\epsilon}^A.
\end{equation}
The displacements $\vec{u}^*$ can be obtained by first converting the stresses $P_{1}^{0,*}$ into forces using Eq.~\ref{eq:force_eq} and then calculating the displacements in the reference network from the forces using Eq.~\ref{eq:u_from_f}:
\begin{equation}
    \label{eq:u_star}
    \vec{u}^* = \left(\boldsymbol{A}^T\boldsymbol{C}^0\boldsymbol{A}\right)^{-1}\boldsymbol{A}^T\vec{P}_1^{0,*}\boldsymbol{S}.
\end{equation}

To understand the non-affine displacement field of jammed disk packings in terms of triangle strains applied to a reference network, we now seek to calculate the eigenstrains that will generate the stresses $\vec{P}_{1}^{0,*}$. The stresses $\vec{P}_{1}^{0,*}$ are in general asymmetric, but the stiffness tensor of the reference network $\boldsymbol{C}^0$ to which the eigenstrains are applied can only generate symmetric stresses. Note that $\vec{P}_{1}^{0,*}$ contains all of the individual triangle stresses $\vec{P}_{1}^{\triangle,*}$: 
\begin{equation}
    \vec{P}_{1}^{0,*} = \begin{bmatrix}
        (\vec{P}_{1}^{\triangle,*})^T & (\vec{P}_{2}^{\triangle,*})^T  & \cdots & (\vec{P}_{2N}^{\triangle,*})^T
    \end{bmatrix}^T.
\end{equation}
To resolve this issue, we can convert the triangle stresses $\vec{P}_{1}^{\triangle,*}$ within $\vec{P}_{1}^{0,*}$ into matrix form $\boldsymbol{P}_{1}^{\triangle,*}$ using Eq.~\ref{eq:mat_to_vec} and decompose $\boldsymbol{P}_{1}^{\triangle,*}$ into symmetric $\mathcal{Q}^{\triangle,*}$ and orthogonal $\mathcal{R}^{\triangle,*}$ matrices~\cite{horn2012matrix}:
\begin{equation}
    \boldsymbol{P}_{1}^{\triangle,*} = \mathcal{R}^{\triangle,*}\mathcal{Q}^{\triangle,*},
\end{equation}
where
\begin{equation}
    \mathcal{Q}^{\triangle,*} = \sqrt{(\boldsymbol{P}^{\triangle,*}_1)^T\boldsymbol{P}^{\triangle,*}_1}
\end{equation}
and
\begin{equation}
    \mathcal{R}^{\triangle,*} = \boldsymbol{P}^{\triangle,*}_{1} (\mathcal{Q}^{\triangle,*})^{-1}.
\end{equation}
Since $\mathcal{Q}^{\triangle,*}$ is symmetric, it can be generated by a strain $\vec{\epsilon}_1^\triangle$ applied to a triangle in the reference network,
\begin{equation}
\label{epsilon1}
    \vec{\mathcal{Q}}^{\triangle,*} = \frac{1}{\mathcal{A}}\boldsymbol{C}^{\triangle,0}\vec{\epsilon}_1^\triangle.
\end{equation}
(Note that ${{\mathcal{Q}}}^{\triangle,*}$ has been converted back into vector form ${\vec {{\mathcal{Q}}}}^{\triangle,*}$.) From Eq.~\ref{epsilon1}, we can solve for the triangle eigenstrains, 
\begin{equation}
    \label{eq:e1}
    \vec{\epsilon}_1^\triangle = \mathcal{A}(\boldsymbol{C}^{\triangle,0})^{-1}\vec{\mathcal{Q}}^{\triangle,*}.
\end{equation}

The expression in Eq.~\ref{eq:e1} allows us to reconstruct the non-affine displacement field of jammed disk packings in terms of localized eigenstrains applied to triangles in the fully connected stress-free reference network. The resulting non-affine displacement field is exact in the limit of small applied strains $\Delta \gamma \rightarrow 0$. Each triangle eigenstrain corresponds to a stiffness mismatch between the jammed disk packing and the reference network. The reference network is nearly homogeneous in its elastic properties and, unlike the corresponding jammed disk packing, it does not possess localized low-frequency modes of the dynamical matrix~\cite{xu2024}. Thus, the reference network is analogous to the reference elastic matrix in the continuum Eshelby inhomogeneity problem. Each triangle eigenstrain (e.g., arising from a missing contact), when applied to the reference network, produces a quadrupolar-like displacement field that is proportional to the magnitude of the eigenstrain. The total non-affine displacement field is thus the sum of the individual quadrupolar-like fields with varying strengths and orientations. 

To examine which triangle defects play a significant role in determining the non-affine displacement field, we first convert the triangle eigenstrain $\vec{\epsilon}_1^\triangle$ to matrix form $\boldsymbol{\epsilon}_1^\triangle$ through Eq.~\ref{eq:mat_to_vec} and then calculate the corresponding von Mises strain
\begin{equation}
    \boldsymbol{\epsilon}_1^\triangle=\boldsymbol{\epsilon}_{1,d}^\triangle+\epsilon_{1,h}^\triangle \boldsymbol{I},
\end{equation}
where the hydrostatic strain $\epsilon_{1,h}=(\epsilon_{1,xx}+\epsilon_{1,yy})/2$. The von Mises strain $\epsilon_{1,v}^\triangle$ is the square root of the second invariant of the deviatoric strain $\boldsymbol{\epsilon}_{1,d}^\triangle$,
\begin{equation}
    \epsilon_{1,v}^\triangle=\sqrt{\frac{1}{2}\textrm{tr} \left((\boldsymbol{\epsilon}_{1,d}^\triangle)^{2}\right)}.
\label{eq:vm}
\end{equation}

\begin{figure}[t]
    \centering
    \includegraphics[width=\linewidth]{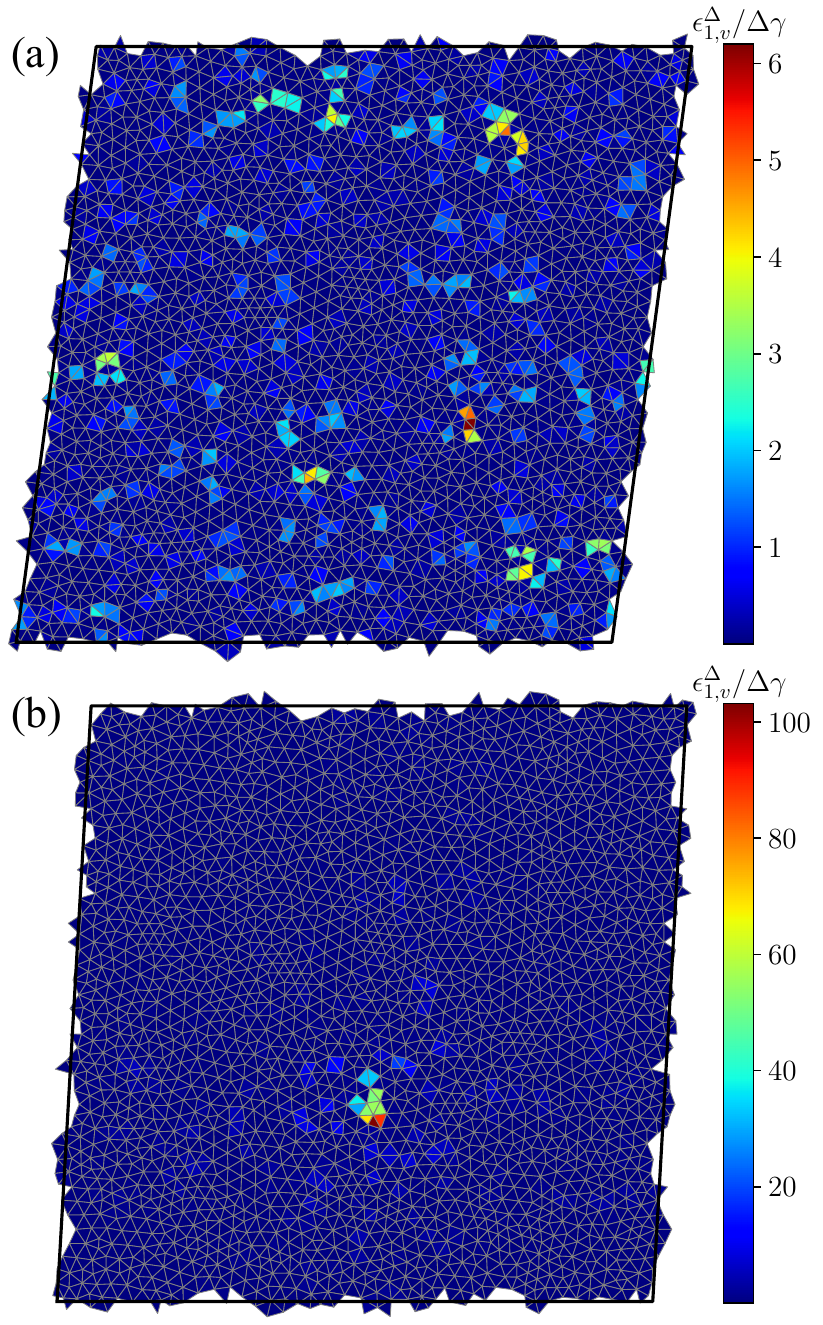}
    \caption{(a) and (b) Spatial map of the von Mises strain $\epsilon_{1,v}^\triangle$ scaled by the shear strain increment $\Delta \gamma$ (increasing from violet to dark red) that generate the non-affine displacement fields in Fig.~\ref{fig:ss} (b) and (c), respectively.}
    \label{fig:mismatch}
\end{figure}

\begin{figure}[t]
    \centering
    \includegraphics[width=\linewidth]{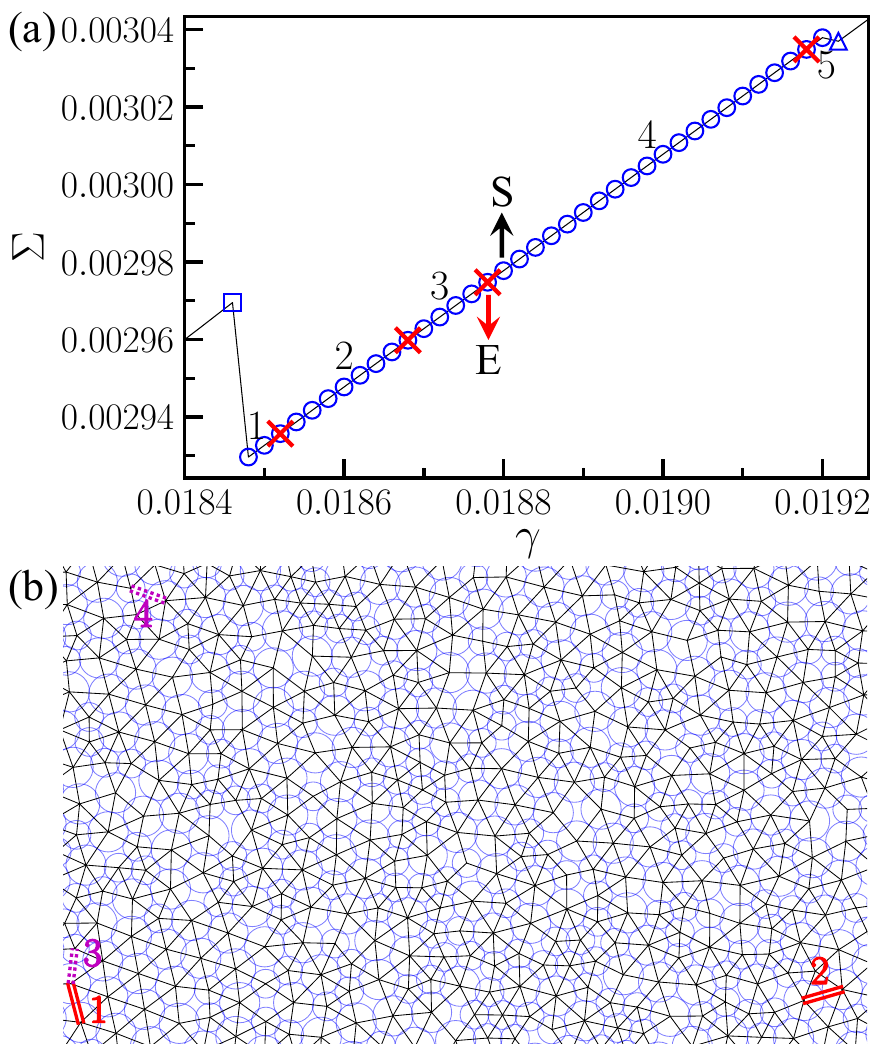}
    \caption{(a) Shear stress $\Sigma$ versus shear strain $\gamma$ (solid line) for a jammed bidisperse disk packing. The circles indicate $\Sigma(\gamma)$ within the central quasi-elastic segment in the figure. The single square and single triangle markers indicate the end of the previous and the start of the next quasi-elastic segment. The central quasi-elastic segment includes $5$ geometrical families (GFs), within which the jammed disk packings at different strains possess the same interparticle contact network. The endpoints of the geometrical families are highlighted by the $\times$ markers. E and S indicate the end of GF $3$ and the start of GF $4$. The transitions between geometrical families within the same quasi-elastic segment involve the breaking or formation of a single contact. (b) Visualization of a portion of the jammed disk packing in (a) at the beginning of the central quasi-elastic segment. Interparticle contacts are depicted as solid lines. The red double solid lines, labeled 1 and 2, indicate the broken contacts that cause the transition from GF $1$ to $2$ and from GF $2$ to $3$. The magenta double dotted lines, labeled 3 and 4, indicate the formation of contacts that cause transitions from GF $3$ to $4$ and from GF $4$ to $5$.  }
    \label{fig:geo}
\end{figure}

In Fig.~\ref{fig:mismatch} (a), we show a spatial map of $\epsilon_{1,v}^\triangle$ for the non-affine displacement field in Fig.~\ref{fig:ss} (b). The von-Mises eigenstrains occur throughout the packing with many large-$\epsilon_{1,v}^\triangle$ regions, which correspond to regions where the magnitude of the non-affine displacement field in Fig.~\ref{fig:ss} (b) is large. In contrast, in Fig.~\ref{fig:mismatch} (b), we observe a small region of triangles with large $\epsilon_{1,v}^\triangle$, which corresponds to the quadrupole-like structure in the non-affine displacement field in Fig.~\ref{fig:ss} (c). The largest $\epsilon_{1,v}^\triangle/\Delta\gamma \gtrsim 100$ in Fig.~\ref{fig:mismatch} (b) is significantly higher than that in Fig.~\ref{fig:mismatch} (a). Using the EIM, we have shown that we can reconstruct the non-affine displacement field using many interacting triangle defects. Previous studies have discussed the difficulty in describing the core regions of quadrupolar-like structures that occur in the non-affine displacement fields of sheared glasses \cite{moriel2020extracting,Albaret2016,rainone2020,Nicolas2018}. Using the method described here, we resolve the core regions of the quadrupole-like structures in the displacement fields in terms of the triangle strains.

\begin{figure}[t]
    \centering
    \includegraphics[width=0.9\linewidth]{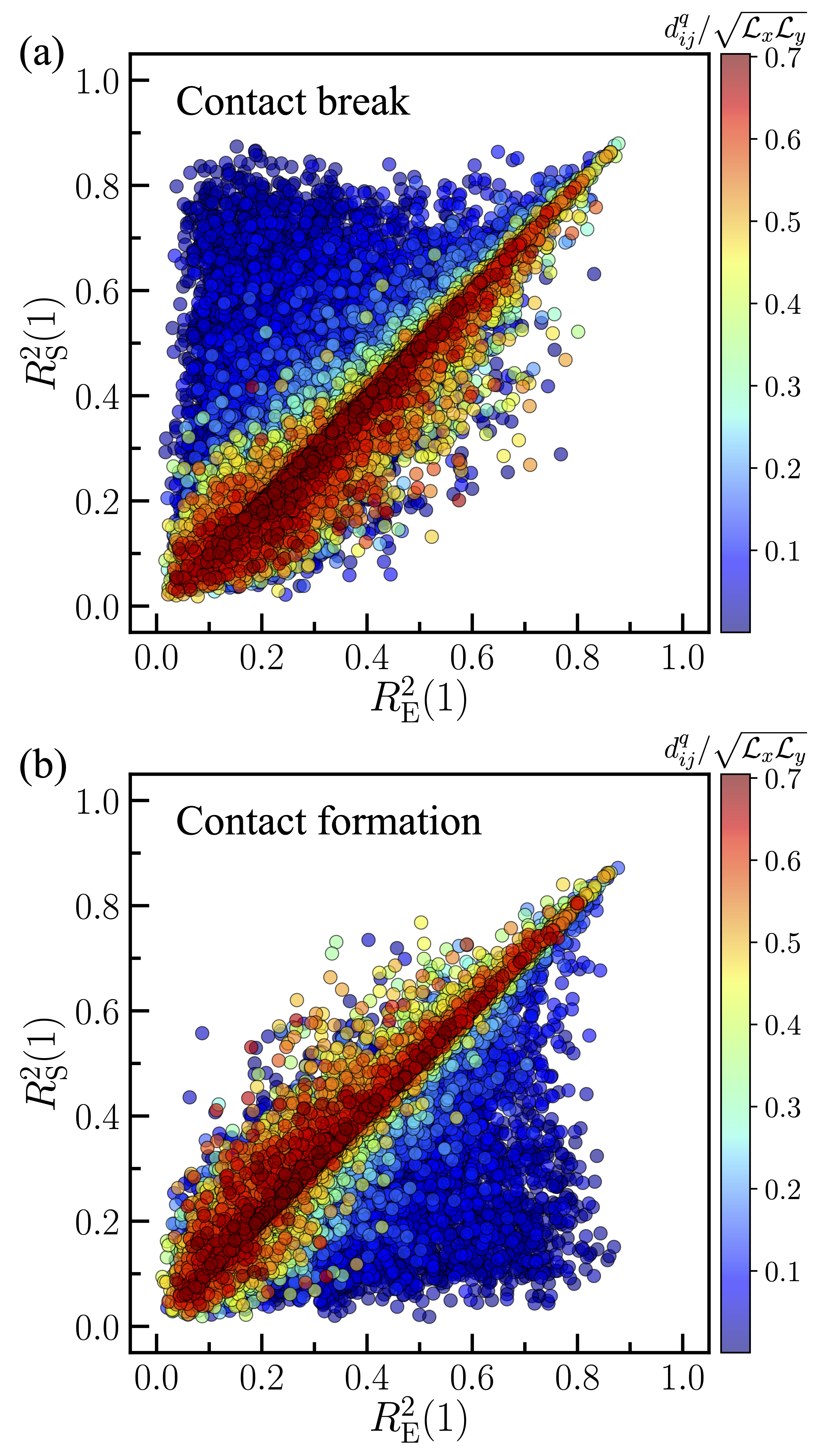}
    \caption{Coefficient of determination $R_{\rm S}^2(1)$ after fitting the non-affine displacement field to Eq.~\ref{fiteq} (for a single effective quadrupole) at the beginning of the next geometrical family plotted as a function of $R_{\rm E}^2(1)$ after fitting the non-affine displacement field at the end of the current geometrical family for two cases: (a) the transition between two geometrical families after a single contact is broken and (b) the transition between two geometrical families after a single contact is formed. The symbols are color-coded based on the distance $d^q_{ij}/\sqrt{\mathcal{L}_x\mathcal{L}_y}$ between the center of the altered contact and the center of the single effective quadrupole, normalized by the geometric mean of edge lengths of the boundary. Panels (a) and (b) include $47,209$ and $35,463$ transitions between geometrical families, respectively.}
    \label{fig:R2R2dist}
\end{figure}

\subsection{Why do quadrupolar displacement fields occur in jammed disk packings undergoing simple shear?}
\label{results:D}

In Sec.~\ref{results:A}, we studied the probability that isolated effective quadrupolar displacement fields occur in jammed disk packings during athermal, simple shear as a function of pressure. We find that triangles with large von Mises eigenstrains (calculated in Sec.~\ref{eigen}) are located near the centers of the effective quadrupoles. However, can we {\it predict} the strain at which single effective quadrupolar displacement fields will occur before the strain is applied? As before, we focus on the non-affine displacement fields that occur during changes in the interparticle contact networks along quasi-elastic stress versus strain segments as shown in Fig.~\ref{fig:geo} (a). Note that the changes in the interparticle contact networks in Fig.~\ref{fig:geo} (b) involve the breaking or formation of a single contact in the small shear strain increment limit ($\Delta \gamma\rightarrow 0$) and these contact changes are reversible~\cite{tuckman2020,lundberg08}.  Below, we will compare the behavior of the non-affine displacement fields at the end (denoted by E) of a given geometric family and the start (denoted by S) of the next geometric family (excluding shear stress drops). In the following, we present results for systems with $N = 2048$ particles at an initial pressure of $p = 0.1$ prior to shear deformation. Comparable findings are observed for other system sizes and pressures where single effective quadrupoles emerge.

In Fig.~\ref{fig:R2R2dist}, we show the results for fits to Eq.~\ref{eq:circEsh} of the displacement field (quantified using the coefficient of determination $R^2$) in response to a single athermal, simple shear strain increment along quasi-elastic stress versus strain segments at the end of a given geometric family and start of the next one. As shown in Fig.~\ref{fig:R2R2dist} (a), changes from disordered non-affine displacement fields to isolated effective quadrupolar displacement fields, where $R_{\rm E}^2(1)$ at the end of a given geometrical family is small and $R_{\rm S}^2(1)$ at the start of the successive geometrical family is large, occur predominantly after a single contact breaks. We find that the single effective quadrupole is centered near the broken contact, which emphasizes the important role of contact breaking in generating quadrupolar structures in the non-affine displacement fields.  In contrast, we show in Fig.~\ref{fig:R2R2dist} (b) that the formation of a new contact near the center of an effective quadrupolar displacement field causes it to dissolve. 

\begin{figure}[t]
    \centering
    \includegraphics[width=0.9\linewidth]{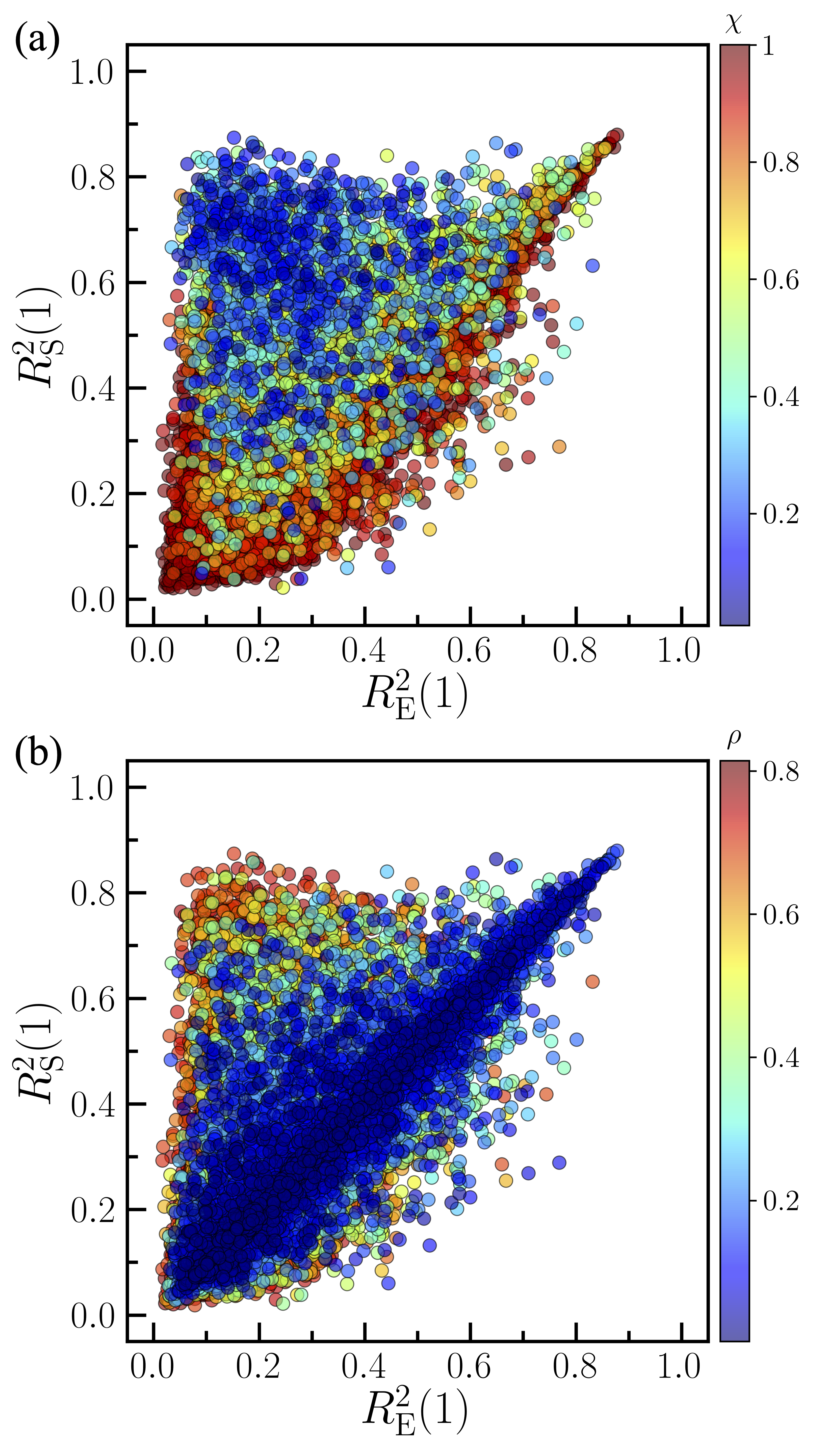}
    \caption{Coefficient of determination $R_{\rm S}^2(1)$ after fitting the non-affine displacement field to Eq.~\ref{fiteq} (for a single effective quadrupole) at the beginning of the next geometrical family plotted as a function of $R_{\rm E}^2(1)$ after fitting the non-affine displacement field at the end of the current geometrical family for transitions between geometrical families with a single contact break (using the same data in Fig.~\ref{fig:R2R2dist} (a)). (a) The symbols are color-coded based on $\chi$, which quantifies the ratio of the lowest non-zero eigenvalue of the dynamical matrix after and before the breaking of a single contact. (b) The symbols are color-coded based on the participation ratio $\rho$ of the eigenvector from the dynamical matrix before the contact is broken that most closely aligns with the eigenvector of the dynamical matrix corresponding to the smallest eigenvalue after the contact is broken.}
    \label{fig:R2R2para}
\end{figure}

\begin{figure*}[t]
    \centering
    \includegraphics[width=\linewidth]{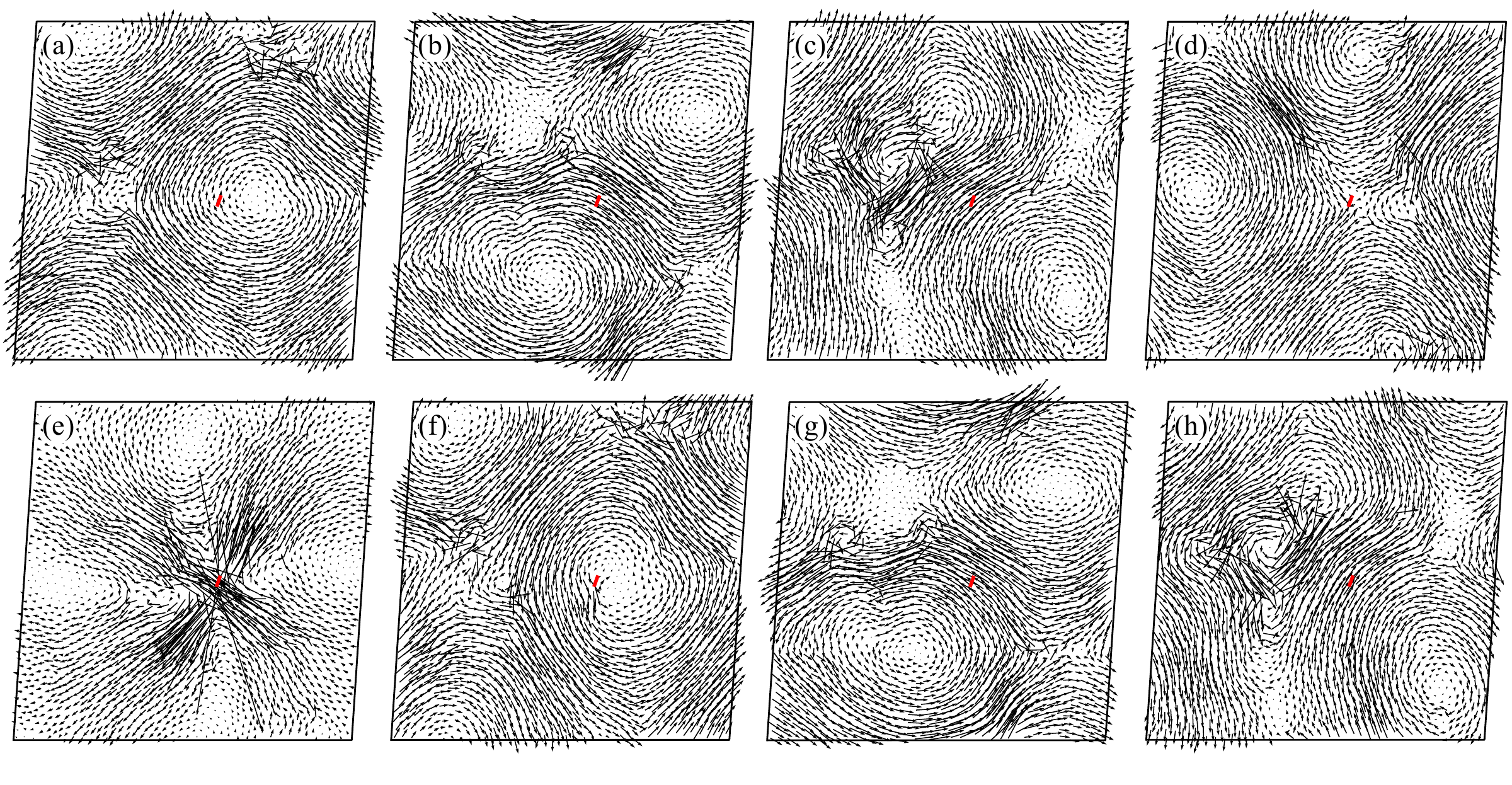}
    \caption{(a)-(d): Spatial maps of the four eigenvectors (corresponding to the four smallest nontrivial eigenvalues of the dynamical matrix) for a jammed disk packing at the end of a geometrical family. The eigenvectors have participation ratios $\rho =0.73$, $0.53$, $0.52$, and $0.69$  and phonon order parameters $O^k=0.89$, $0.86$, $0.69$, and $0.89$, respectively. (e)-(h): Spatial maps of the four eigenvectors (corresponding to the four smallest nontrivial eigenvalues of the dynamical matrix) of the jammed packing at the beginning of the new geometrical family following a single shear strain increment applied to the jammed packing in panels (a)-(d). The transition between the two geometrical families involves the breaking of a single contact (with $\chi=0.126$) given by the thick red line in each panel.}
    \label{fig:eig}
\end{figure*}

While contact breaking is necessary for the formation of single effective quadrupoles, it is not always sufficient for quadrupole formation, as illustrated in Fig.~\ref{fig:R2R2dist} (a). To investigate this question, we show in Fig.~\ref{fig:R2R2para} (a) the variation of $R^2_S(1)$ versus $R^2_E(1)$ following a contact break, color-coded by $\chi$, which quantifies the ratio of the lowest non-zero eigenvalue of the dynamical matrix after and before a contact break. (The mathematical definition of $\chi$ is provided in Appendix~\ref{appendix:D}.) We find that the formation of an isolated effective quadrupole is correlated with relatively small values of $\chi$, which indicates that the broken contact should be in close alignment with the eigenvectors associated with small dynamical matrix eigenvalues of the jammed disk packing before the contact break. 

\begin{figure}[h]
    \centering
    \includegraphics[width=0.9\linewidth]{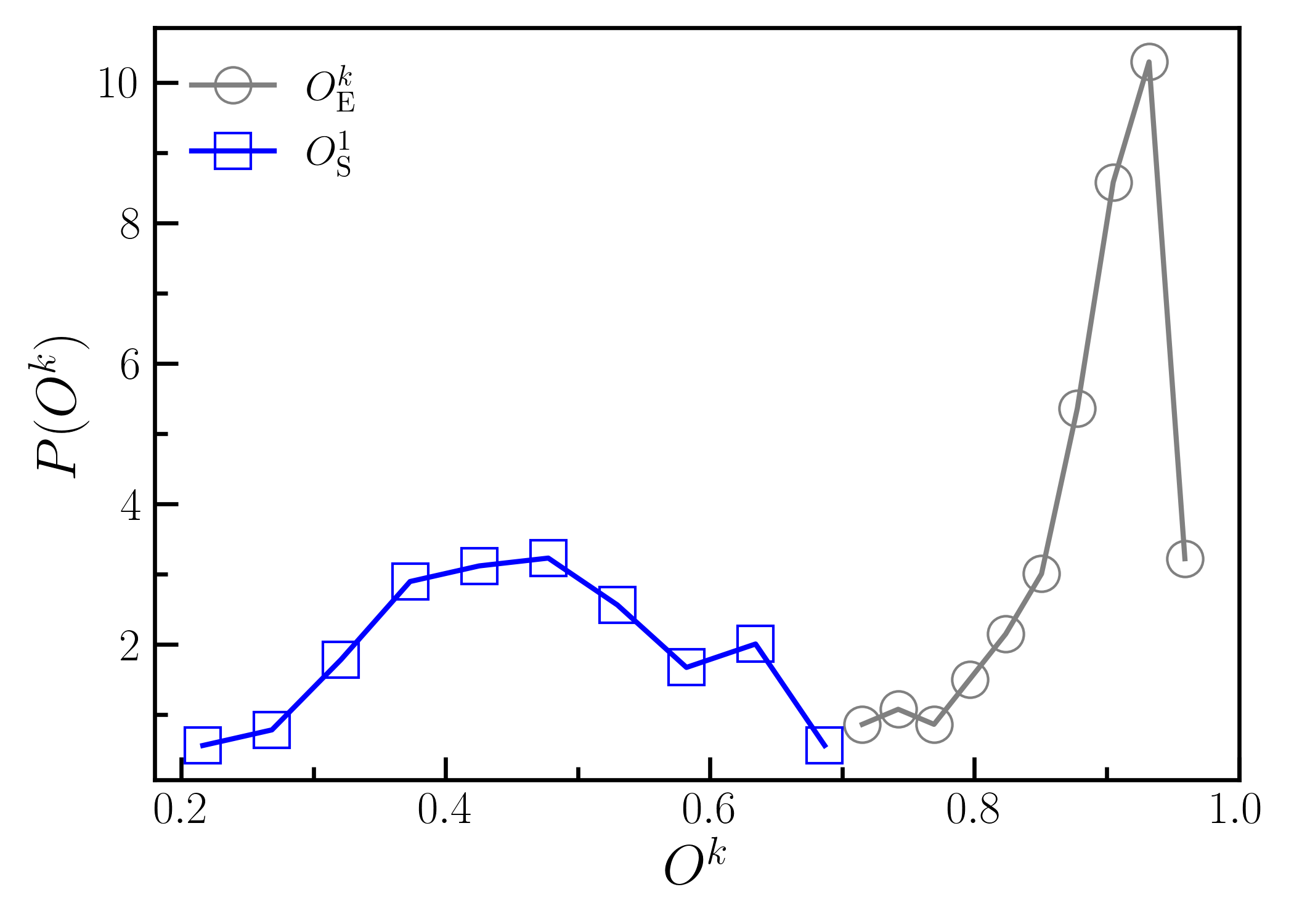}
    \caption{ Probability distribution $P(O^k)$ of the phonon order parameter for eigenmodes at transitions between geometrical families with $R_{\rm E}^2(1)<0.2$ and $R_{\rm S}^2(1)>0.7$ in Fig.~\ref{fig:R2R2para}. $P(O^k_{\rm E})$ corresponds to the distribution for the vibrational mode before contact breaking that aligns most closely with the lowest-frequency mode after the contact breaking. $P(O^1_{\rm S})$ shows the corresponding distribution of the phonon order parameter of the lowest-frequency mode after contact breaking.}
    \label{fig:pdf_phonon}
\end{figure}

We also investigated the correlation of $R^2(1)$ for the non-affine displacement fields at contact breaks with the participation ratio of the eigenvectors of the dynamical matrix,
\begin{equation}
\rho(\vec{e}_{k}) = \frac{\left(\sum_{i=1}^N \|\vec{e}_{k, i}\|^2\right)^2}{N\sum_{i=1}^N \|\vec{e}_{k, i}\|^4},
\end{equation}
where $\vec{e}_{k}$ is the $k$th normalized eigenvector of the dynamical matrix corresponding to eigenvalue $\lambda_k$, and $\vec{e}_{k, i}$ is the contribution to $\vec{e}_{k}$ from particle $i$. In Fig.~\ref{fig:R2R2para} (b), we plot the participation ratio $\rho$ of the eigenvector before the contact break $\vec{e}_{k, \rm E}$ that most closely aligns with the eigenvector corresponding to the smallest nonzero eigenvalue $\vec{e}_{1, \rm S}$ (after contact breaking), i.e., we identify $k$ such that $|\vec{e}_{k, \rm E}\cdot \vec{e}_{1, \rm S}|$ is maximized. Note that data points in Fig.~\ref{fig:R2R2para} (b) with lower values of $\rho$ are placed on top of the other data points. The distribution of participation ratios in Fig.~\ref{fig:R2R2para} (b) $P(\rho)$ is shown in Fig.~\ref{fig:pdf_PR} in Appendix~\ref{appendix:E}. We find that single effective quadrupoles form when $\vec{e}_{k, \rm E}$ has a relatively large value of $\rho$, i.e., when the eigenvector is extended rather than highly localized. These results emphasize that isolated effective quadrupoles form during athermal, simple shear when (1) the interparticle contact that breaks is aligned with the low-frequency modes of the jammed disk packing before the contact break and (2) the low-frequency modes of the jammed disk packing before the contact break are extended rather than highly localized. 
 We also quantified the phonon order parameter $0 \le O^k \le 1$ for each eigenmode, where $O^k=1$ for a phonon-like mode and $O^k=0$ for a mode that is non-phonon-like~\cite{mizuno2017continuum}. Our analysis focuses on cases in which breaking a single contact gives rise to a single effective quadrupole, corresponding to transitions between geometrical families with $R_{\rm E}^2(1)<0.2$ and $R_{\rm S}^2(1)>0.7$ in Fig.~\ref{fig:R2R2para}. In Fig.~\ref{fig:pdf_phonon}, we show the probability distributions of the phonon order parameter for (1) the mode before a contact break that best aligns with the lowest-frequency mode after the contact break ($P(O^k_{\rm E})$) and (2) the lowest frequency mode after the 
contact break ($P(O^1_{\rm S})$). The values of $O^k_{\rm E}$ (before contact breaking) are typically close to 1, with a distribution that is sharply peaked near $0.9$. This result indicates that these extended modes are phonon-like. After the contact break, however, the lowest-frequency mode becomes quasi-localized, with $O^1_{\rm S} \sim 0.45$.

Figure~\ref{fig:eig} shows a spatial map of the four eigenvectors that correspond to the four lowest nontrivial eigenvalues across a geometrical family transition involving a single contact break. Prior to the contact break, the non-affine displacement field is disordered without a single or small number of effective quadrupoles (i.e. $R^2_{\rm E}(1)=0.175$). In contrast, following the contact break, a single effective quadrupole forms in the displacement field with $R^2_{\rm S}(1)=0.841$. The newly broken contact has $\chi \approx 0.126$ with the eigenvectors of the dynamical matrix from the jammed packing before the broken contact. As shown in Figs.~\ref{fig:eig} (a)-(d), the low-frequency modes before the broken contact are extended with $\rho>0.5$ and phonon-like with $O^k\geq0.69$. After the contact break (Fig.~\ref{fig:eig} (e)-(h)), the eigenvector $\vec{{e}}_1$ corresponding to the lowest nontrivial eigenvalue becomes highly localized at the location of the broken contact, with a significantly reduced participation ratio of $\rho=0.054$. This low-frequency eigenmode is closely aligned with the corresponding non-affine displacement field $\vec{u}$, i.e., $|\vec{{e}}_1\cdot \vec{u}|/\|\vec{u}\| \approx 0.998$.

We have shown that a newly broken contact is the primary trigger that gives rise to a single effective quadrupole. How do existing missing contacts relative to the fully connected Delaunay network (i.e., not the newly broken contact) affect the formation of quadrupoles? In the previous section, we demonstrated that Delaunay triangles with large von Mises eigenstrains occur frequently near the centers of isolated quadrupoles. Further, the large-eigenstrain triangles typically have at least one missing contact, which suggests that the missing contacts (other than the one that triggers the quadrupole) may play an important role in quadrupole formation. However, jammed disk packings contain many missing contacts relative to the fully connected network. For example, in the jammed disk packing in Fig.~\ref{fig:mismatch}, there are $N_m = 556$ missing contacts relative to the fully connected network with $3N = 6144$ interparticle contacts. Thus, which of the many missing contacts associated with the large-eigenstrain triangles are important for triggering and dissolving quadrupolar displacement fields?

To address this question, we develop a protocol to identify the \emph{key missing contacts} responsible for quadrupole formation. Specifically, for a jammed disk packing with $N_m$ missing contacts, we ``heal'' each missing contact between disk $i$ and $j$ one at a time by setting $\sigma_{ij}$ in the interparticle potential in Eq.~\ref{potential} to their current separation.  This method increases the stiffness of the two Delaunay triangles sharing the healed contact without changing the potential energy or interparticle forces. 
After healing a given missing contact, the modified packing with $N_m - 1$ missing contacts undergoes a single athermal, quasistatic simple shear strain increment. The resulting non-affine displacement field is then fit to a single effective quadrupole, which yields $R^2_{ij}(1)$. We repeat this process for each of the $N_m-1$ other missing contacts.

\begin{figure}[t]
    \centering
    \includegraphics[width=0.9\linewidth]{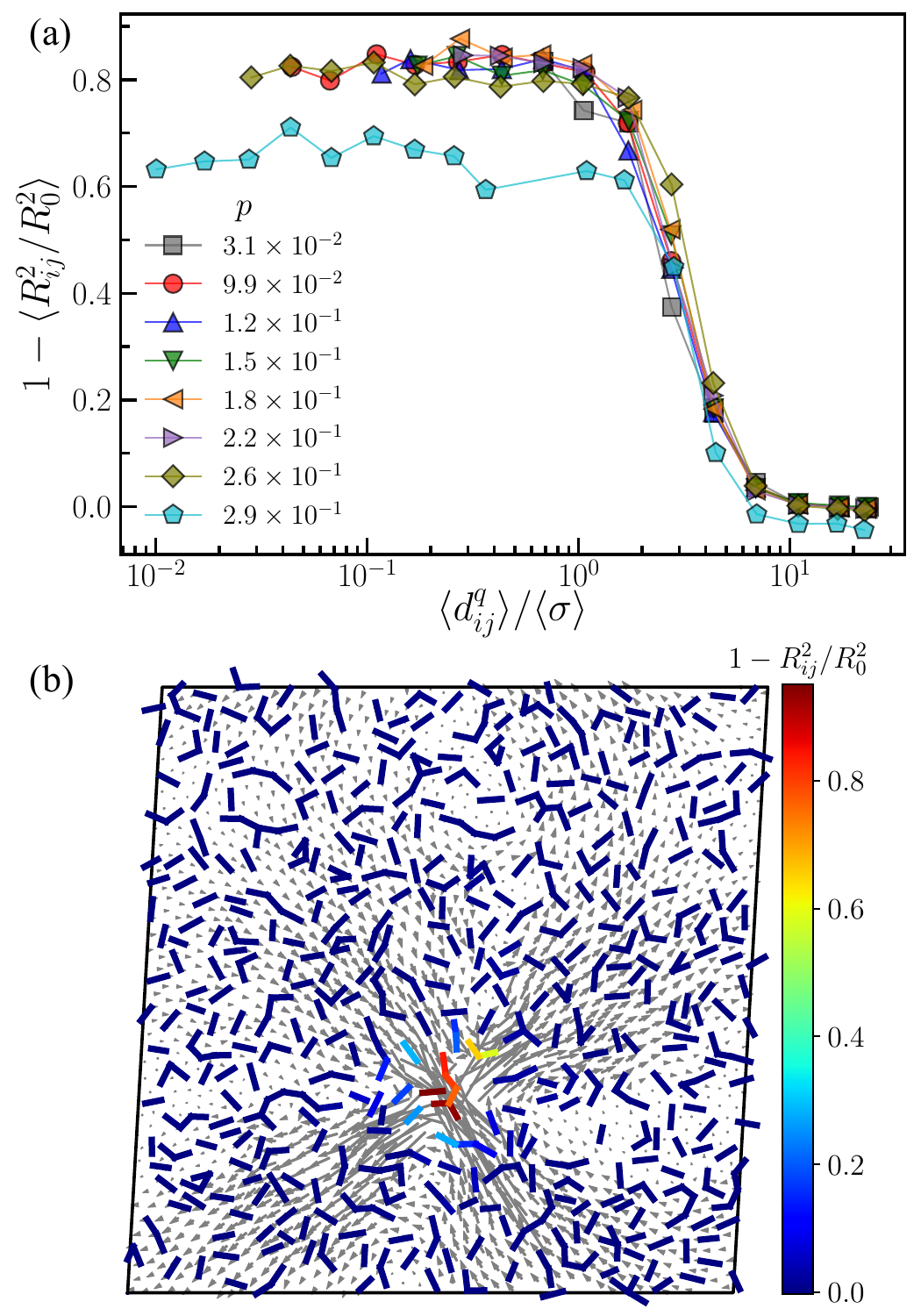}
    \caption{(a) Relative change in the coefficient of determination $1-R^2_{ij}/R^2_{0}$ after healing the missing contact between disks $i$ and $j$ plotted versus the scaled distance $d^q_{ij}/\langle \sigma\rangle$, where $d^q_{ij}$ is the distance from the center of the original quadrupole to that of the healed contact and $\langle\sigma\rangle$ is the average disk diameter. $R^2_{0}>0.7$ is the coefficient of determination from the fit of the original non-affine displacement field before healing to a single effective quadrupole. (b) Illustration of the missing contacts (thick solid lines) in a jammed disk packing with a single effective quadrupolar non-affine displacement field (arrows). Each missing contact between disks $i$ and $j$ is colored by the corresponding value of $1-R^2_{ij}/R^2_{0}$ after it is healed.}
    \label{fig:keymiss}
\end{figure}

In Fig.~\ref{fig:keymiss} (a), we show the relative change in the coefficient of determination after the healing of each missing contact, $1-R^2_{ij}/R^2_{0}$, where $R^2_{0}>0.7$ is the coefficient of determination from the fit of the displacement field before healing the contacts. The results in Fig.~\ref{fig:keymiss} (a) reveal that healing missing contacts located far from the quadrupole center does not affect the non-affine displacement field, since $1-R^2_{ij}/R^2_{0}$ is small at large distances $d^q_{ij}/\langle \sigma\rangle$ from the original quadrupole. In contrast, healing a missing contact within approximately $3\langle\sigma\rangle$ of the quadrupole causes it to dissolve. These contacts are thus classified as key missing contacts.
In Fig.~\ref{fig:keymiss} (b), we display a map of $1-R^2_{ij}/R^2_{0}$ for all missing contacts in an example jammed disk packing, which emphasizes that the important missing contacts are clustered near the center of the quadrupole. The localization of the missing contacts also correlates with the regions of large eigenstrains.

\section{Conclusions and future directions}
\label{summary}
In this work, we investigated the non-affine displacement fields of 2D jammed packings of frictionless disks interacting via purely repulsive linear spring potentials undergoing athermal, quasistatic simple shear. We first demonstrated that the probability that the non-affine displacement fields possess one or two effective quadrupoles increases with pressure. While isolated effective quadrupolar structures are rare at low pressures, the probability increases sharply for pressure $p \gtrsim 0.2$, reaching $33\%$ for one and $76\%$ for two effective quadrupoles at the maximum pressures studied here. 

We then showed that quadrupolar displacement fields, particularly in jammed packings with a single missing contact compared to the fully connected Delaunay-triangulated network, closely resemble the solutions of the Eshelby inclusion and inhomogeneity problems in continuum materials. Each missing contact in jammed disk packings introduces a pronounced local stiffness difference compared to the reference Delaunay network, which can give rise to a quadrupolar displacement field provided that the shear direction is neither parallel nor perpendicular to the missing bond. Similar to Eshelby inclusions in continuum materials, the orientation of the quadrupoles aligns with the direction of the missing bond. 
Given the large number of missing contacts in jammed disk packings compared to the reference network, we reformulated Eshelby’s equivalent inclusion method (EIM) for jammed disk packings with multiple interacting triangular inclusions. The resulting superposition of triangle eigenstrains (applied to the reference network) yields an exact representation of the non-affine displacement field for jammed disk packings undergoing athermal, quasistatic simple shear in the $\Delta \gamma \rightarrow 0$ limit. This new framework emphasizes that the particle-scale origins of quadrupolar displacement fields are triangle stiffness mismatches between the jammed disk packing and the reference network.

We also investigated the necessary conditions for quadrupolar displacement fields to form in jammed disk packings during athermal, quasistatic simple shear. Our results show that the formation of an isolated effective quadrupole requires: (1) relatively extended low-frequency vibrational modes for the jammed packing before the applied deformation and (2) the breaking of a contact that is aligned with the low-frequency vibrational modes. The quadrupole that forms is centered on the newly broken contact, but numerous important existing missing contacts (relative to the reference network) near the effective quadrupole can affect its stability.  For example, we showed that if we heal an existing missing contact within $\sim 3\langle\sigma\rangle$ of the quadrupole center, the effective quadrupole dissolves. 

These results open several important questions for future research. 
First, we showed that each triangle stiffness mismatch pertaining to a missing contact is associated with a quadrupolar displacement field, while the interaction of multiple triangle mismatches can lead to either disordered non-affine displacement fields lacking coherent structures or coherent displacement fields with a few isolated effective quadrupoles. We also found that contact breaking is a necessary condition for the formation of an isolated quadrupole. However, the sufficient conditions for quadrupole formation have not yet been determined. Second, a single effective quadrupole, spanning most of the system, is often observed in the $N=2048$ systems considered in the present studies. We anticipate that there will be a well-defined number density of effective quadrupoles emerging simultaneously that depends on pressure $p$ in the large-system limit. Third, the EIM for jammed disk packings developed here can be extended to amorphous solids with long-range interactions. In this case, we will likely need to modify the reference network to include second, third, or further neighboring particles. The EIM can also be extended to three dimensions. In 3D systems, the low-frequency vibrational modes are typically more localized than those in two dimensions~\cite{mizuno2017continuum,kapteijns_2018}. This stronger localization makes it more difficult to detect the quadrupolar symmetry that characterizes the nonaffine displacement field near a broken contact. We can investigate whether the tetrahedra obtained by Delaunay tetrahedralization of the packing can serve as the reference networks for jammed sphere packings in 3D and capture quadrupolar displacement fields triggered by the stiffness mismatches between tetrahedra. Using the EIM in 2D and 3D, we will also explore the mechanisms that control shear band formation in amorphous solids undergoing shear deformation~\cite{Sopu2017}. 
One hypothesis is that localized plastic events propagate along shear planes by activating a network of triangle- or tetrahedral units with stiffness mismatches, which drive strain localization and eventually catastrophic failure. Alternatively, shear bands may arise from pre-existing alignment of such mismatches with the shear direction. By tracking the motion of Eshelby-like defects during relaxation for large shear stress drops, we can determine whether shear bands are initiated through the progressive triggering of Eshelby-like mismatches or whether large shear bands form all at once from inherent structural anisotropy within the material. Further, we will probe the role of static friction in the formation, motion, and dissolution of quadrupoles in jammed granular materials. Addressing these questions will deepen our understanding of structural defect-mediated deformation in amorphous solids, such as granular media.

\begin{acknowledgments}
We acknowledge support from NSF Grant No. DGE-2244310 (E.P.W. and C.S.O.), NSF Grant No. CMMI-1901959
(W.J., A.D., U.D.S., and C.S.O.), and ARO Grant No. W911NF-23-1-0032 (W.J. and C.S.O.). This work was also supported by the High Performance Computing facilities operated by Yale's Center for Research Computing.
\end{acknowledgments}

\section*{DATA AVAILABILITY}
The data that support the findings of this article are openly available~\cite{data}, embargo periods may apply.

\appendix

\section{Eshelby Inclusion Problem for Continuum Materials}
\label{appendix:A}

\begin{figure}
    \centering
    \includegraphics[width=\linewidth]{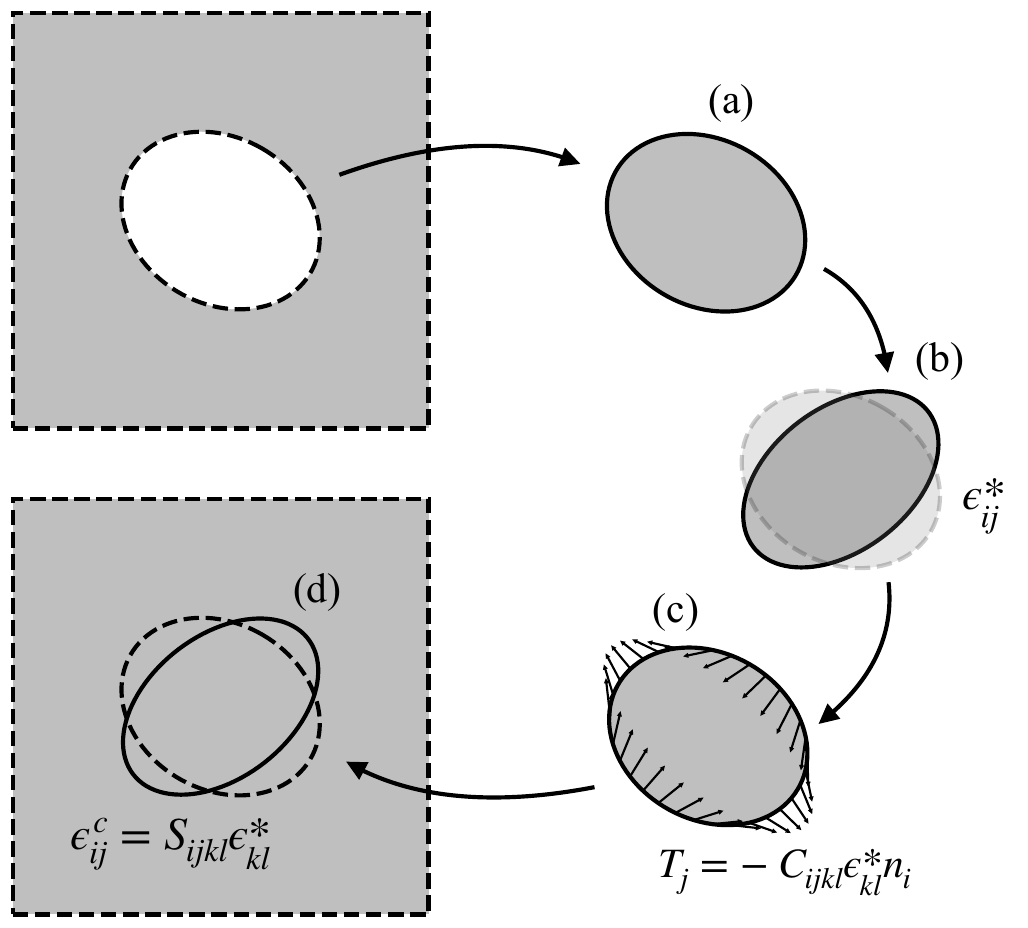}
    \caption{The solution to the Eshelby inclusion problem in continuum materials can be obtained using a series of cutting, straining, and welding operations: (a) First an elliptical inclusion is cut out of an infinite solid matrix; (b) It then undergoes a plastic eigenstrain $\epsilon^*_{ij}$ with an original shape given by the dashed line; (c) A traction $T_j=-C_{ijkl}\epsilon^*_{kl}n_i$ ($\vec{n}$ is the outward normal unit vector) is applied to deform the inclusion into its original shape, where $C_{ijkl}$ is the stiffness matrix of the material; (d) The inclusion is reinserted into the matrix. Equilibrating the stresses results in a final deformed shape (indicated by the solid line) with strain $\epsilon^c_{ij}=S_{ijkl}\epsilon^*_{kl}$, where $S_{ijkl}$ is the Eshelby tensor that depends on the elastic properties of the material and inclusion geometry. The original shape is indicated by the dashed line.}
    \label{fig:continuum_inclusion}
\end{figure}

The Eshelby inclusion problem in continuum materials assumes that an elliptical inclusion within an infinite solid undergoes a spontaneous plastic eigenstrain $\epsilon^*_{ij}$. Due to the surrounding matrix material, stresses will emerge in both the inclusion and matrix, resulting in the final strain, or constrained strain $\epsilon^c_{ij}$, of the inclusion to be different from that of the eigenstrain. The Eshelby inclusion problem can be solved using the series of operations in Fig.~\ref{fig:continuum_inclusion}. 
First, an elliptical inclusion is cut out of an infinite solid matrix and allowed to deform into a new shape where the strain is given by the eigenstrain $\epsilon^*_{ij}$.  A traction is then applied to the surface of the inclusion to strain it back into its original shape. The inclusion is reinserted and welded back to the matrix material. The inclusion is then allowed to relax into a new deformed shape with strain $\epsilon^c_{ij}$. 
The eigenstrain and constrained strain are related through $\epsilon^c_{ij}$ = $S_{ijkl}\epsilon_{kl}^*$, where $S_{ijkl}$ is the Eshelby tensor that is a function of the stiffness of the material and geometry of the inclusion. The process to solve the Eshelby inclusion problem for particulate systems, such as jammed disk packings, is shown in Fig.~\ref{fig:discrete_inclusion}. For jammed disk packings, we first remove a triangle from the reference network, strain the triangle, and reset the equilibrium lengths of the bonds to achieve zero stress in the triangle. Forces are then applied to the three vertices of the triangle to deform it into its original shape. The triangle is reinserted into the network and the system is allowed to relax.

\begin{figure*}
    \centering
    \includegraphics[width=\linewidth]{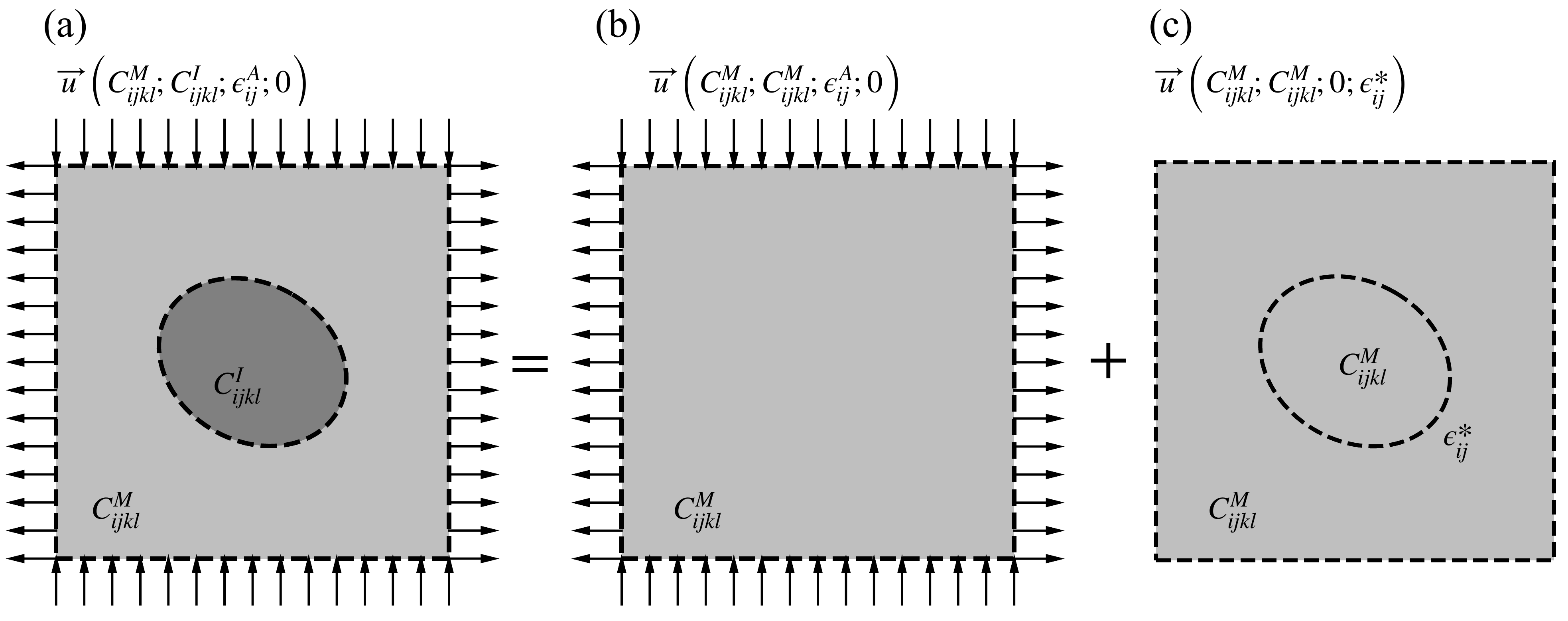}
    \caption{A schematic illustrating the displacement field solution to the Eshelby inhomogeneity problem for an elliptical inclusion in an infinite solid, where $C_{ijkl}^I$ and $C_{ijkl}^M$ are the inclusion and matrix stiffness tensors. (a) The displacement field $\vec{u}\left(C_{ijkl}^M; C_{ijkl}^I; \epsilon_{ij}^A; 0\right)$ can be decomposed into (b) the displacement field for an affine pure shear strain applied to the matrix material $\vec{u}\left(C_{ijkl}^M; C_{ijkl}^M; \epsilon_{ij}^A; 0\right)$ plus (c) the displacement field for an eigenstrain (Eq.~\ref{eq:EIM}) applied to the inclusion with $C_{ijkl}^I = C_{ijkl}^M$.}
    \label{fig:continuum_inhomogeneity}
\end{figure*}

The solution to the Eshelby inclusion problem can be used to solve the Eshelby inhomogeneity problem, i.e., an affine strain $\epsilon^A_{ij}$ is applied to a material with an elliptical inclusion that has a stiffness mismatch with the matrix material. As shown in Fig.~\ref{fig:continuum_inhomogeneity}, the displacement field solution for a given inclusion geometry $\vec{u}\left(C_{ijkl}^M;C_{ijkl}^I; \epsilon_{ij}^A;\epsilon_{ij}^*\right)$, which is a function of the stiffness tensors of the matrix $C_{ijkl}^M$ and inclusion $C_{ijkl}^I$ and the applied affine strain $\epsilon_{ij}^A$, is the sum of two contributions:  the displacement field for the affine strain applied to the matrix material and the displacement field for an eigenstrain $\epsilon_{ij}^*$ applied to the inclusion with $C_{ijkl}^I=C_{ijkl}^M$. The eigenstrain for this decomposition can be obtained using Eq.~\ref{eq:EIM}.

As shown in Fig.~\ref{fig:discrete_inhomogeneity}, a similar approach can be carried out to solve the Eshelby inhomogeneity problem in particulate systems, such as jammed disk packings. For jammed disk packings, the corresponding matrix material is the Delaunay-triangulated reference network. In Fig.~\ref{fig:discrete_inhomogeneity}, the inclusion is a triangle with bonds that have different spring constants from the surrounding network. The displacement field solution for an affine strain applied to the network with the triangle defect can be written as the sum of two contributions: the displacement field of a reference spring network without a defect undergoing the same affine strain plus the displacement field from eigenstrains applied to the triangles in the reference network that had the mismatching spring constants in original network.

\section{Components of the Triangle Stiffness Matrix}
\label{appendix:B}
In this appendix, we calculate the components of the triangle stiffness matrix $\boldsymbol{C}^\triangle$ (Eq.~\ref{eq:triangle_C_mat}) by expressing the triangle potential energy $U^\triangle$ in terms of the deformation gradient (Eq.~\ref{eq:U_wrt_F})
\begin{equation}
    \boldsymbol{F}^\triangle = \begin{bmatrix}
        F_{xx} & F_{xy} \\ 
        F_{yx} & F_{yy}
    \end{bmatrix},
\end{equation}
taking the second derivatives of $U^\triangle$ with respect to $\boldsymbol{F}^\triangle$, and evaluating the second derivatives at $\boldsymbol{F}^\triangle=\boldsymbol{I}$. The components of the triangle stiffness matrix are 
\begin{equation}
\boldsymbol{C}^\triangle = \begin{bmatrix}
        c_{xxxx} & c_{xxyy} & c_{xxxy} & c_{xxyx} \\
        c_{yyxx} & c_{yyyy} & c_{yyxy} & c_{yyyx} \\
        c_{xyxx} & c_{xyyy} & c_{xyxy} & c_{xyyx} \\
        c_{yxxx} & c_{yxyy} & c_{yxxy} & c_{yxyx}
    \end{bmatrix},
\end{equation}
with
\begin{equation}
    c_{xxxx} = \frac{\partial^2U}{\partial F_{xx}\partial F_{xx}} = \sum_{\alpha} \frac{L_{x \alpha}^2k_{\alpha}}{L_{\alpha}^3}\left(L_{\alpha}L_{x \alpha}^2 +b_{\alpha}L_{y \alpha}^2\right),
\end{equation}

\begin{equation}
    \label{eq:cxxyy}
    c_{xxyy} = \frac{\partial^2U}{\partial F_{xx}\partial F_{yy}} = \sum_{\alpha} \frac{L_{x \alpha}^2L_{y \alpha}^2k_{\alpha}}{L_{\alpha}^3}\left(L_{\alpha} -b_{\alpha}\right),
\end{equation}

\begin{equation}
    \label{eq:cxxxy}
    c_{xxxy} = \frac{\partial^2U}{\partial F_{xx}\partial F_{xy}} =  \sum_{\alpha} \frac{L_{x \alpha}L_{y \alpha}k_{\alpha}}{L_{\alpha}^3}\left(L_{\alpha}L_{x \alpha}^2 +b_{\alpha}L_{y \alpha}^2\right),
\end{equation}

\begin{equation}
    c_{xxyx} = \frac{\partial^2U}{\partial F_{xx}\partial F_{yx}} \sum_{\alpha} \frac{L_{x \alpha}^3L_{y \alpha}k_{\alpha}}{L_{\alpha}^3}\left(L_{\alpha} - b_{\alpha}\right),
\end{equation}

\begin{equation}
    c_{yyyy} = \frac{\partial^2U}{\partial F_{yy}\partial F_{yy}} = \sum_{\alpha} \frac{L_{y \alpha}^2k_\alpha}{L_\alpha^3}\left(L_\alpha L_{y \alpha}^2 +b_\alpha L_{x \alpha}^2\right),
\end{equation}

\begin{equation}
    c_{yyxy} = \frac{\partial^2U}{\partial F_{yy}\partial F_{xy}} = \sum_{\alpha} \frac{L_{x \alpha}L_{y \alpha}^3k_\alpha}{L_\alpha^3}\left(L_\alpha -b_\alpha\right),
\end{equation}

\begin{equation}
    c_{yyyx} =\frac{\partial^2U}{\partial F_{yy}\partial F_{yx}} =  \sum_{\alpha} \frac{L_{x \alpha}L_{y \alpha}k_\alpha}{L_\alpha^3}\left(L_\alpha L_{y \alpha}^2 +b_\alpha L_{x \alpha}^2\right),
\end{equation}

\begin{equation}
    c_{xyxy} = \frac{\partial^2U}{\partial F_{xy}\partial F_{xy}}= \sum_{\alpha} \frac{L_{y \alpha}^2k_\alpha}{L_\alpha^3}\left(L_\alpha L_{x \alpha}^2 +b_\alpha L_{y \alpha}^2\right),
\end{equation}

\begin{equation}
    c_{xyyx} = \frac{\partial^2U}{\partial F_{xy}\partial F_{yx}}= \sum_{\alpha} \frac{L_{x \alpha}^2L_{y \alpha}^2k_\alpha}{L_\alpha^3}\left(L_\alpha -b_\alpha\right),
\end{equation}
and
\begin{equation}
    c_{yxyx} = \frac{\partial^2U}{\partial F_{yx}\partial F_{yx}} =  \sum_{\alpha} \frac{L_{x \alpha}^2k_\alpha}{L_\alpha^3}\left(L_\alpha L_{y \alpha}^2 +b_\alpha L_{x \alpha}^2\right).
\end{equation}

From Eqs.~\ref{eq:cxxyy} and~\ref{eq:cxxxy}, we can see that since the two successive derivatives can be interchanged
\begin{equation}
    c_{yyxx} = c_{xxyy}
\end{equation}
and
\begin{equation}
    c_{xyxx} = c_{xxxy}.
\end{equation}
For similar reasons,  
\begin{equation}
    c_{yxxx} = c_{xxyx},
\end{equation}
\begin{equation}
    c_{xyyy} = c_{yyxy},
\end{equation}
\begin{equation}
    c_{yxyy} = c_{yyyx},
\end{equation}
and 
\begin{equation}
    c_{yxxy} = c_{xyyx}.
\end{equation}

\section{Construction of the global stiffness matrices}
\label{appendix:C}

To apply the EIM to jammed disk packings, we must define the gradient matrix $\boldsymbol{A}$ and stiffness matrix $\boldsymbol{C}$ for the jammed packing.  We first relate the displacements to the strains for the entire system by expressing the $8N\times2N$ $\boldsymbol{A}$ matrix in terms of the  $4\times6$ $\boldsymbol{A}^\triangle$ matrices for each of the $2N$ triangles. We can then calculate the global first Piola-Kirchhoff stresses $\vec{P}_1$ (to first order in the triangle strains) by defining the $8N\times8N$ global stiffness matrix $\boldsymbol{C}$ in terms of the $4 \times 4$ triangle stiffness matrices $\boldsymbol{C}^\triangle$.

To construct the global $8N\times2N$ $\boldsymbol{A}$ gradient matrix from the local $4\times6$ triangle $\boldsymbol{A}^\triangle$ gradient matrices, we multiply a $8N\times12N$ matrix with all of the triangle $\boldsymbol{A}^\triangle$ gradient matrices along the diagonal (with $4\times6$ zero matrices elsewhere) by a $12N\times2N$ matrix with $2\times2$ identity matrices in the columns that correspond to the nodal numbering of the triangle vertices in the global system (and $2\times2$ zero matrices elsewhere):
\begin{widetext}
\begin{equation}
    \label{eq:construct_A}
    \boldsymbol{A} = 
    \begin{bmatrix}
    \boldsymbol{A}^\triangle_1 & \boldsymbol{0} & \cdots & \cdots&\boldsymbol{0}\\
    \boldsymbol{0}  & \ddots & \cdots & \cdots & \vdots\\
    \vdots & \vdots & \boldsymbol{A}^\triangle_\beta & \cdots & \vdots \\
    \vdots & \vdots & \vdots &  \ddots & \boldsymbol{0}\\
    \boldsymbol{0} & \cdots  &\cdots & \boldsymbol{0} & \boldsymbol{A}^\triangle_{2N}
    \end{bmatrix}
     \begin{blockarray}{cccccccccc}
    1 & 2 & \cdots & i & j & \cdots & m & \cdots & N \\
    \begin{block}{[ccccccccc]c}
      \boldsymbol{0} & \boldsymbol{0} & \cdots & \boldsymbol{0} & \boldsymbol{0} & \cdots & \boldsymbol{0} & \cdots & \boldsymbol{0}& 1 \\
       \boldsymbol{0} & \boldsymbol{0} & \cdots & \boldsymbol{0} & \boldsymbol{0} & \cdots & \boldsymbol{0} & \cdots & \boldsymbol{0}& 2 \\
       \boldsymbol{0} & \boldsymbol{0} & \cdots & \boldsymbol{0} & \boldsymbol{0} & \cdots & \boldsymbol{0} & \cdots & \boldsymbol{0}& 3 \\
      \vdots & \vdots & \vdots &\vdots & \vdots &\vdots & \vdots &\vdots &\vdots &\vdots \\
      \boldsymbol{0} & \boldsymbol{0} & \cdots & \boldsymbol{I} & \boldsymbol{0} & \cdots &\boldsymbol{0}& \cdots & \boldsymbol{0}& 3\beta-2 \\
      \boldsymbol{0} & \boldsymbol{0} & \cdots & \boldsymbol{0} & \boldsymbol{I} & \cdots &\boldsymbol{0}& \cdots & \boldsymbol{0}& 3\beta-1 \\
      \boldsymbol{0} & \boldsymbol{0} & \cdots & \boldsymbol{0} & \boldsymbol{0} & \cdots &\boldsymbol{I}& \cdots & \boldsymbol{0}& 3\beta \\
      \vdots & \vdots & \vdots &\vdots & \vdots &\vdots & \vdots &\vdots & \vdots &\vdots \\
      \boldsymbol{0} & \boldsymbol{0} & \cdots & \boldsymbol{0} & \boldsymbol{0} & \cdots & \boldsymbol{0} & \cdots & \boldsymbol{0}& 6N-2\\
      \boldsymbol{0} & \boldsymbol{0} & \cdots & \boldsymbol{0} & \boldsymbol{0} & \cdots & \boldsymbol{0} & \cdots & \boldsymbol{0}& 6N-1\\
      \boldsymbol{0} & \boldsymbol{0} & \cdots & \boldsymbol{0} & \boldsymbol{0} & \cdots & \boldsymbol{0} & \cdots & \boldsymbol{0}& 6N,\\
    \end{block}
    \end{blockarray}
\end{equation}
\end{widetext}
where $\boldsymbol{A}^\triangle_\beta$ for triangle $\beta$ has nodes $i$, $j$, and $m$. 
To construct the global stiffness matrix $\boldsymbol{C}$, we place the triangle $4\times4$ stiffness matrices $\boldsymbol{C}^\triangle$ along the diagonal of a $8N\times8N$ matrix: 
\begin{equation}
\boldsymbol{C} = \begin{bmatrix}
   \boldsymbol{C}_{1}^\triangle & \boldsymbol{0} & \cdots & \boldsymbol{0} \\
   \boldsymbol{0} & \boldsymbol{C}_{2}^\triangle & \cdots & \boldsymbol{0} \\
   \vdots  & \vdots  & \ddots & \vdots  \\
   \boldsymbol{0} & \boldsymbol{0} & \cdots & \boldsymbol{C}_{2N}^\triangle
 \end{bmatrix}.
\end{equation}
The global first Piola-Kirchhoff stress $\vec{P}_1$ is an $8N\times1$ column vector defined as
\begin{equation}
    \vec{P}_1 = \boldsymbol{C}\vec{\epsilon}\boldsymbol{S}^{-1},
\end{equation}
where the $8N\times8N$ matrix
\begin{equation}
    \boldsymbol{S} =  \begin{bmatrix}
   \mathcal{I}\mathcal{A}_{1} & \boldsymbol{0} & \cdots & \boldsymbol{0} \\
   \boldsymbol{0} & \mathcal{I}\mathcal{A}_{2} & \cdots & \boldsymbol{0} \\
   \vdots  & \vdots  & \ddots & \vdots  \\
   \boldsymbol{0} & \boldsymbol{0} & \cdots & \mathcal{I}\mathcal{A}_{2N}
\end{bmatrix}
\end{equation}
$\mathcal{I}$ is a $4\times4$ identity matrix, and ${\cal A}_{\beta}$ is the area of triangle $\beta$. $\vec{\epsilon}$ is the $8N\times1$ global strain column vector consisting of all of the individual $4\times1$ triangle strains $\vec{\epsilon}^\triangle$,
\begin{equation}
    \label{eq:global_strain}
    \vec{\epsilon} = \begin{bmatrix}
        (\vec{\epsilon}_{1}^{\triangle})^T & (\vec{\epsilon}_{2}^{\triangle})^T  & \cdots & (\vec{\epsilon}_{2N}^{\triangle})^T
    \end{bmatrix}^T.
\end{equation}

\section{Variation of the lowest non-zero eigenvalue of the dynamical matrix}
\label{appendix:D}

In this appendix, we characterize the low-frequency eigenvalues of the dynamical matrix in jammed disk packing before and after a single contact break. The dynamical matrix is defined as 
\begin{equation}
    {M}_{i\alpha j\beta} = \frac{\partial^2 U}{\partial r_{i\alpha}\partial r_{j\beta}},
    \label{eq:dm}
\end{equation}
where $U=\sum_{i>j} U_{ij}(r_{ij})$ is the total potential energy and $r_{i\alpha}$ is the $\alpha$-component of particle position $\vec{r}_i=[x_i, y_i]^T$. 
The contribution to the dynamical matrix $\boldsymbol{M}$ from the interaction potential $U_{ij}$ between disks $i$ and $j$ is
\begin{equation}
    {D}^{(ij)}_{p\alpha q\beta} = \left.\frac{\partial^2 U_{ij}(r_{ij})}{\partial r_{p\alpha}\partial r_{q\beta}}\right\vert_{p,q \in \{i,j\}},
\end{equation}
whose elements satisfy the symmetry relations ${D}^{(ij)}_{j\alpha j\beta} = {D}^{(ij)}_{i\alpha i\beta} = -{D}^{(ij)}_{j\alpha i\beta} = -{D}^{(ij)}_{i\alpha j\beta}$. 

\begin{figure}[t]
    \centering
    \includegraphics[width=0.9\linewidth]{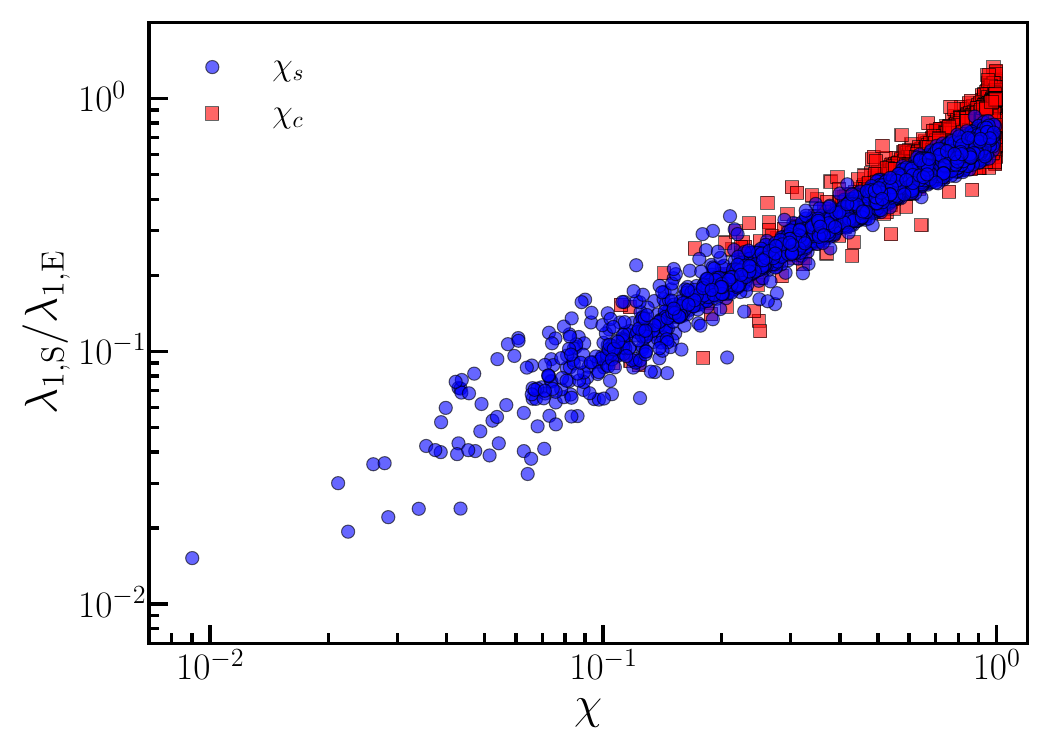}
    \caption{Ratio $\lambda_{1, \rm S}/\lambda_{1, \rm E}$ of the lowest non-zero eigenvalue from the dynamical matrix of a jammed disk packing after the broken contact to that before the broken contact plotted as a function of $\chi_s$ in Eq.~\ref{eq:estlbda} (blue circles) and $\chi_c$ in Eq.~\ref{eq:estlbdb} (red squares).}
    \label{fig:lbdchi}
\end{figure}

For transitions between geometrical families involving a single contact break, the contact with the smallest overlap is the one that breaks. Hence, the contribution of the pair $(i,j)$ at the strain just prior to contact breaking satisfies
\begin{equation}
\label{extra}
\begin{aligned}
\boldsymbol{D}^{(ij)}_{i i} &\quad=
    \begin{bmatrix}
    1-\frac{\sigma_{ij}}{r_{ij}}\sin^2\theta_{ij} & \frac{\sigma_{ij}}{r_{ij}}\sin\theta_{ij}\cos\theta_{ij} \\
    \frac{\sigma_{ij}}{r_{ij}}\sin\theta_{ij}\cos\theta_{ij} & 1-\frac{\sigma_{ij}}{r_{ij}}\cos^2\theta_{ij}
    \end{bmatrix},
\end{aligned}
\end{equation}
where $\theta_{ij}$ is the angle between $\vec{r}_{ij}=\vec{r}_{i}-\vec{r}_{j}$ and the $x$-axis. In the limit of small particle overlap ($r_{ij} \rightarrow \sigma_{ij}$) near contact breaking, Eq.~\ref{extra} becomes
\begin{equation}
\begin{aligned}
\boldsymbol{D}^{(ij)}_{i i} &\quad=
    \begin{bmatrix}
    \cos^2\theta_{ij} & \sin\theta_{ij}\cos\theta_{ij} \\
    \sin\theta_{ij}\cos\theta_{ij} & \sin^2\theta_{ij}
    \end{bmatrix}.
\end{aligned}
\end{equation}
$\boldsymbol{D}^{(ij)}$ is a rank-one matrix, i.e., $\boldsymbol{D}^{(ij)} =\vec{v}\vec{v}^T$, where $\vec{v}$ is a column vector that indicates the orientation of the broken bond with four elements $\cos\theta_{ij}$, $\sin\theta_{ij}$, $-\cos\theta_{ij}$, and $-\sin\theta_{ij}$ for the $ix$, $iy$, $jx$, and $jy$ indexes, respectively. (The other elements in the dynamical matrix are zero.) The lowest non-zero eigenvalue $\tilde{\lambda}_1$ of the dynamical matrix $\tildeb{\boldsymbol{M}}=\boldsymbol{M}-\boldsymbol{D}^{(ij)}$ after the contact break can be obtained by solving 
\begin{equation}
\label{extra2}
\begin{aligned}
    \det\left( \tildeb{\boldsymbol{M}} - \tilde{\lambda}_1 I\right) &= \det\left( \boldsymbol{M} - \tilde{\lambda}_1 I\right) \left[ 1-\vec{v}^T\left(\boldsymbol{M} - \tilde{\lambda}_1 I\right)^{-1}\vec{v} \right]\\
    &= 0.
\end{aligned}
\end{equation}
Eq.~\ref{extra2} can be rewritten as 
\begin{equation}
\label{extra3}
    1-\sum_{k=1}^{N_e}\frac{\left(\vec{e}_k^T \vec{v}\right)^2}{\lambda_k-\tilde{\lambda}_1}=0, 
\end{equation}
where $\vec{e}_k$ is the normalized eigenvector of $\boldsymbol{M}$ with eigenvalue $\lambda_k$. The eigenvalues of the dynamical matrix can be sorted as $\lambda_1<\lambda_2<\dots<\lambda_{N_e}$, where $N_e$ is the number of non-zero eigenvalues. 

For $\tilde{\lambda}_1 \ll \lambda_1$, the $\frac{1}{\lambda_k-\tilde{\lambda}_1}$ factor in Eq.~\ref{extra3} can be approximated as
\begin{equation}
\frac{1}{\lambda_k-\tilde{\lambda}_1}\approx\frac{1}{\lambda_k}+\frac{\tilde{\lambda}_1}{\lambda_k^2}, \quad k=1, 2, \dots, N_e,
\end{equation}
and substituted into Eq.~\ref{extra3} to estimate the ratio
\begin{equation}
\frac{\tilde{\lambda}_1}{\lambda_1} \approx \chi_s = \frac{1-\sum_{k=1}^{N_e}\frac{\left(\vec{e}_k^T \vec{v}\right)^2}{\lambda_k}}{\lambda_1\sum_{k=1}^{N_e}\frac{\left(\vec{e}_k^T \vec{v}\right)^2}{\lambda_k^2}},
\label{eq:estlbda}
\end{equation}
which depends on the alignment of $\vec{v}$ with all eigenvectors of the dynamical matrix.

To obtain an improved estimate of $\tilde{\lambda}_1$ when it is close to $\lambda_1$, we keep the $\frac{1}{\lambda_1-\tilde{\lambda}_1}$ term and use a zeroth-order approximation for $k>1$:
\begin{equation}
\frac{1}{\lambda_k-\tilde{\lambda}_1}\approx\frac{1}{\lambda_k}.
\end{equation}
Hence, Eq.~\ref{extra3} becomes
\begin{equation}
    1-\frac{\left(\vec{e}_1^T \vec{v}\right)^2}{\lambda_1-\tilde{\lambda}_1}-\sum_{k=2}^{N_e}\frac{\left(\vec{e}_k^T \vec{v}\right)^2}{\lambda_k}\approx0, 
\end{equation}
which yields
\begin{equation}
\frac{\tilde{\lambda}_1}{\lambda_1} \approx \chi_c = \frac{1-\sum_{k=1}^{N_e}\frac{\left(\vec{e}_k^T \vec{v}\right)^2}{\lambda_k}}{1-\sum_{k=2}^{N_e}\frac{\left(\vec{e}_k^T \vec{v}\right)^2}{\lambda_k}},
\label{eq:estlbdb}
\end{equation}
where $\chi_c$ is dominated by the alignment of $\vec{v}$ with $\vec{e}_1$.

We define the parameter $\chi$ as the smaller of the estimated values of $\tilde{\lambda}_1/\lambda_1$ from Eqs.~\ref{eq:estlbda} and \ref{eq:estlbdb}, i.e.,
\begin{equation}
    \chi = \min \left( \chi_s, \chi_c\right).
\end{equation}
The ratio $\lambda_{1,S}/\lambda_{1,E}$ of the smallest nontrivial eigenvalues of the dynamical matrix for jammed disk packings before and after a single contact break (where $\lambda_{1,E}$ and $\lambda_{1,S}$ are from the jammed packings before and after the contact break, respectively) is plotted as a function of $\chi$ in Fig.~\ref{fig:lbdchi}. The strong correlation between $\lambda_{1,S}/\lambda_{1,E}$ and $\chi$ suggests that $\chi$ can be used to predict changes in the smallest nontrivial eigenvalues of the dynamical matrix.

\begin{figure}[t]
    \centering
    \includegraphics[width=0.8\linewidth]{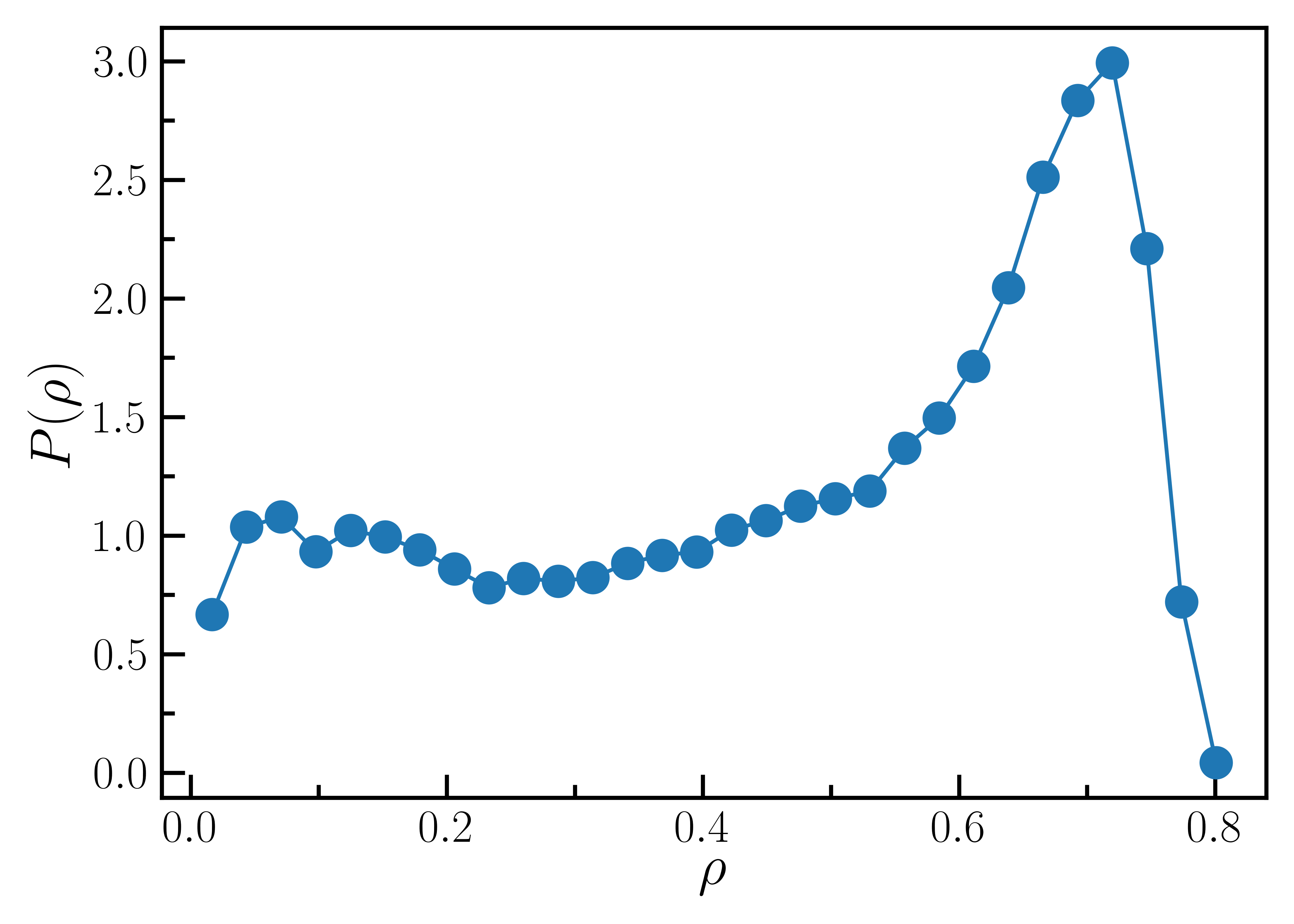}
    \caption{ Probability distribution $P(\rho)$ of the participation ratio $\rho$ of the vibrational modes along a geometrical family before contact breaking as shown in Fig.~\ref{fig:R2R2para} (b).  }
    \label{fig:pdf_PR}
\end{figure}

\begin{figure}[t]
    \centering
    \includegraphics[width=0.825\linewidth]{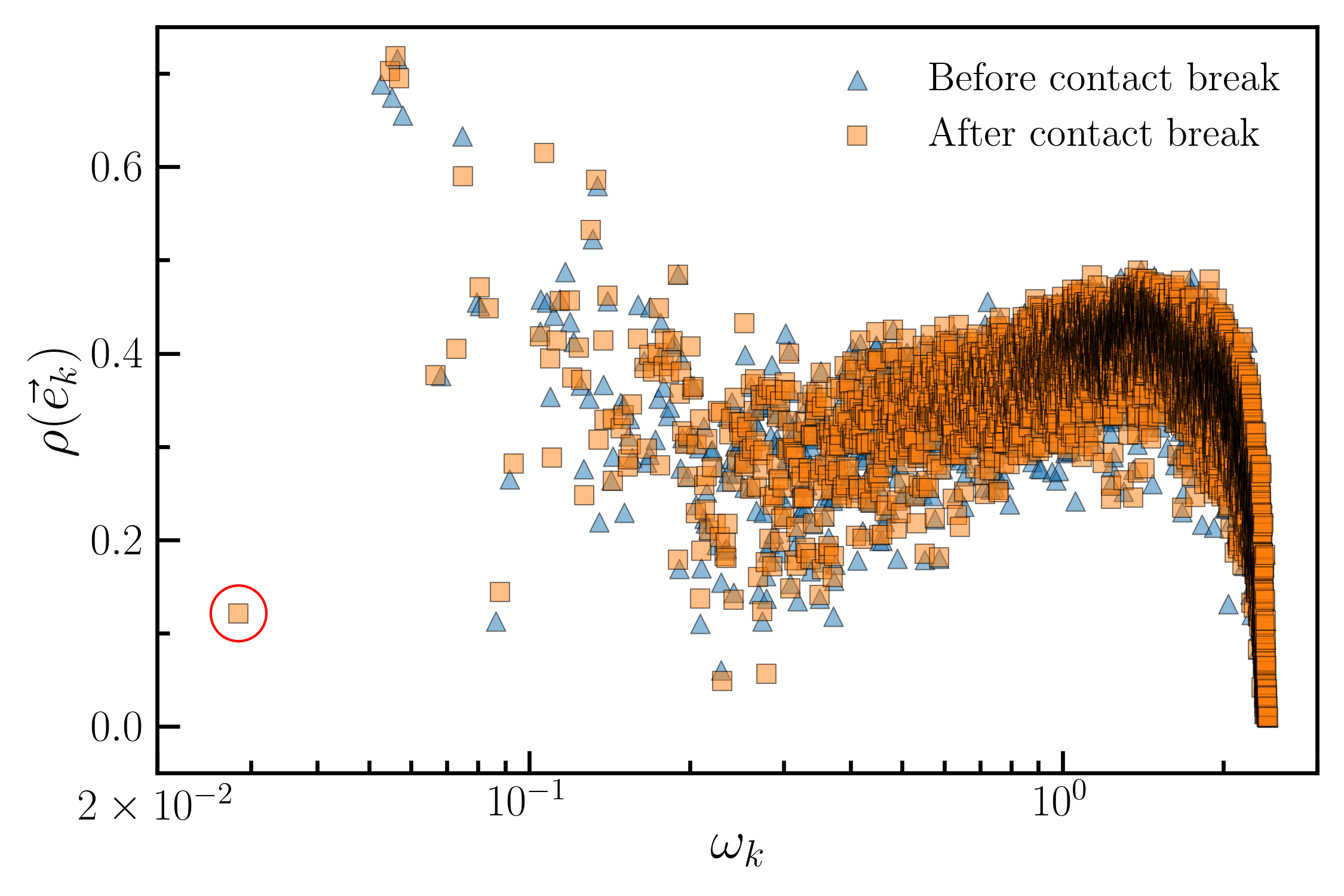}
    \caption{ Participation ratio $\rho$ of the eigenvector $\vec{e}_{k}$ plotted as a function of the corresponding vibrational frequency $\omega_{k}=\sqrt{\lambda_k/m}$ for packings near the transition between two geometrical families ($R^2_{\rm E}(1)=0.102$ and $R^2_{\rm S}(1)=0.835$). Triangles and squares correspond to the packings immediately before and after a single contact break, respectively.  }
    \label{fig:PR_freq}
\end{figure}

\section{Participation ratio of eigenvectors of the dynamical matrix before and after contact breaking}
\label{appendix:E}
The apparent abundance of low-$\rho$ modes in Fig.~\ref{fig:R2R2para} (b) is caused by the fact that data points with lower participation ratio $\rho$ are placed on top of the other data points. Our intention is to highlight cases in which the breaking of a single contact gives rise to a single effective quadrupole. By displaying the low-$\rho$ points on top, we emphasize that the upper-left region of Fig.~\ref{fig:R2R2para} (b), which corresponds to the formation of isolated effective quadrupoles, possesses large values of $\rho$. 
Figure~\ref{fig:pdf_PR} shows the distribution of the participation ratio $P(\rho)$ in Fig.~\ref{fig:R2R2para} (b). The distribution is peaked at $\rho \approx 0.7$. In general, contact breaking leads to more localized low-frequency modes, as $\rho(\vec{e}_{1,\rm S})$ is often smaller than $\rho(\vec{e}_{1,\rm E})$. This result is illustrated in Fig.~\ref{fig:PR_freq}, which tracks the participation ratio during the formation of an isolated effective quadrupole. Before the contact break, the non-affine displacement field has $R^2_{\rm E}(1) \approx 0.102$. After the break, an isolated effective quadrupole emerges with $R^2_{\rm S}(1) \approx 0.835$. The contact break gives rise to a lower-frequency mode at $\omega_k \approx 3 \times 10^{-2}$, with a significantly reduced participation ratio, $\rho(\vec{e}_{1,\rm S}) \approx 0.12$ (highlighted by a circle in Fig.~\ref{fig:PR_freq}), compared to $\rho(\vec{e}_{1,\rm E}) \approx 0.7$ before the contact break.

\end{document}